\documentclass[pdflatex,sn-mathphys-num]{sn-jnl}
\usepackage{graphicx}
\usepackage[export]{adjustbox}
\graphicspath{{figures/}{./}}
\setkeys{Gin}{max width=\linewidth,max totalheight=0.82\textheight,keepaspectratio}
\usepackage{forloop}
\usepackage{etoolbox}
\usepackage{calc}
\usepackage{wrapfig,lipsum,booktabs}
\usepackage{xtab}
\usepackage[dvipsnames]{xcolor}
\usepackage[normalem]{ulem}
\usepackage{placeins}
\usepackage{amsmath,amsfonts,amssymb,amsthm,epsfig,epstopdf,url,array}
\usepackage[retainorgcmds]{IEEEtrantools}
\usepackage{makeidx,lscape}
\usepackage{pict2e}
\usepackage{upgreek}
\usepackage[braket,qm]{qcircuit}
\usepackage{appendix}
\usepackage{bbold}
\usepackage{bbm}
\usepackage{comment}

\title[Diagnosing Device Performance in Rydberg-Ladder Gauge Simulators]{Diagnosing Device Performance in Rydberg-Ladder Gauge Simulators with Cumulative Probabilities and Filtered Mutual Information}

\author*[1]{\fnm{Avi} \sur{Kaufman}}
\author[1,2]{\fnm{Muhammad} \sur{Asaduzzaman}}
\author[1]{\fnm{Zane} \sur{Ozzello}}
\author[1]{\fnm{Blake} \sur{Senseman}}
\author[1]{\fnm{James} \sur{Corona}}
\author[1]{\fnm{Yannick} \sur{Meurice}}

\affil*[1]{\orgdiv{Department of Physics and Astronomy}, \orgname{The University of Iowa}, \orgaddress{\city{Iowa City}, \state{IA}, \postcode{52242}, \country{USA}}}
\affil[2]{\orgdiv{Department of Physics}, \orgname{North Carolina State University}, \orgaddress{\city{Raleigh}, \state{NC}, \postcode{27607}, \country{USA}}}

\abstract{We study bitstring measurements from the publicly available Aquila Rydberg-atom platform using a two-leg ladder that encodes a truncated lattice gauge model as a practical benchmark that can be directly implemented and simulated on current hardware. Our goal is diagnostic: we analyze how errors propagate into bitstring probability distributions and downstream information measures, focusing on ladders with 6, 8, and 10 rungs and $\mathcal{O}(10^3)$ shots. We introduce cumulative probability distributions as a compact way to compare Aquila data with high-accuracy density matrix renormalization group (DMRG) and exact references, and we use optimally filtered mutual information primarily as a robust device-data diagnostic rather than a direct entanglement estimator. By isolating finite sampling, sorting fidelity, adiabatic ramp-up, Rabi-frequency ramp-down, and readout errors, we find that readout mitigation performs well in controlled DMRG tests. Applying the same procedure on hardware shows accuracy limitations for the leading probabilities estimation, indicating that readout errors are not dominant and that residual error is instead driven by imperfect state preparation.}


\begin{document}
\maketitle
\def\beq{\begin{equation}}
\def\enq{\end{equation}}
\def\nq{n_q}
\def\nmax{n_{\mathrm{max}}}
\def\phix{\hat{\phi} _{\bf x}}
\def\nq{n_q}
\def\ah{\hat{a}}
\def\ahd{\hat{a}^\dagger}
\def\pn{P_{\nmax-1}}
\def\har{\hat{H}^{\mathrm{har.}}}
\def\hanh{\hat{H}^{\mathrm{anh.}}}
\def\hpf{\hat{\phi}^4}
\def\np{\mathcal{N}_p}
\def\pt{\tilde{p}}
\def\npc{\mathcal{N}(p)}
\def\dpc{\frac{\Delta p}{2}}
\def\nsh{N_{sh}}

\section{Introduction}
There has been a lot of recent interest in quantum simulation for models used in condensed matter, particle, and nuclear physics \cite{Bloch:2008zzb,Altman:2019vbv,Cloet:2019wre,Alexeev:2020xrq,Klco:2021lap,Bauer:2022hpo,DiMeglio:2023nsa}. 
An area of focus is the real-time evolution for lattice gauge theory models \cite{Tagliacozzo:2012vg,Zohar:2015hwa,Banuls:2019bmf,Meurice:2020pxc,Halimeh:2022mct,Halimeh:2023lid}.

In this context, the controlled manipulation of small quantum systems has opened the possibility of analog calculations of the real-time evolution of lattice models. In the following, we focus on arrays of  Rydberg atoms \cite{Bernien2017Dynamics, Keesling2019Kibble, Labuhn2016RydIsing,Leseleuc2019topo,Ebadi2021_256,Pascal2021AF,Shaw_2024,Choi:2021npc} used to build quantum simulators for lattice gauge theory models 
\cite{Zhang:2018ufj,Surace:2019dtp,Notarnicola:2019wzb,Celi:2019lqy,Meurice:2021pvj, Fromholz:2022ymy,Gonzalez-Cuadra:2022hxt,
Heitritter:2022jik,Bauer:2022hpo,Halimeh:2023lid} and more specifically on a ladder geometry \cite{Meurice:2021pvj,Zhang:2023agx,floating} that we use here as a practical benchmark that can be easily simulated with current hardware. Our focus in this paper is not to introduce a new gauge-theory mapping, but to use this established ladder setup as a controlled benchmark for diagnosing hardware performance from measured bitstring distributions. 

It should be emphasized that for particle physics there is no physical lattice and one is mostly interested 
in studying the continuum limits  where the correlation lengths cover many lattice spacings. 
An important quantity to study phase transitions and the critical behavior is the entanglement entropy \cite{amico2008entanglement,Eisert:2008ur,Abanin_2019,Cirac:2020obd,Ghosh:2015iwa,VanAcoleyen:2015ccp,Banuls:2017ena,Knaute:2024wfh, Kharzeev:2017qzs, PhysRevD.98.054007,Zhang:2021hra,Beane:2018oxh,Robin:2020aeh, PhysRevLett.90.227902,PhysRevLett.92.096402,Calabrese:2005zw,Ryu:2006bv}. The von Neumann entanglement entropy $S^{vN}_A$, with $A$ as one half of the ladder simulator considered in this article, was studied in \cite{floating} as a function of the experimentally tunable parameters (lattice spacing, Rabi frequency and detuning). Its graphical representation shows a rich set of lobes and transition regions.

Recent cold atom experiments \cite{Islam2015,Kaufman2016} have  measured the second-order R\'enyi entropy, $S_2$, by preparing twin copies of the ground state and applying the beamsplitter operation. One can measure 
the number of particles modulo 2 at each site $j$ of a given copy ( $n_j^{copy}$) \cite{Islam2015}, and use the result from
\cite{PhysRevLett.109.020505}
\begin{equation}
\exp (-S_2) = {\rm Tr}(\rho _A^2)= 
\langle (-1)^{\sum_{j\in A} n^{copy}_j}\rangle \ ,
\label{eq:parities}
\end{equation}
\noindent
to calculate $S_2$. 
The reduced density matrix $\hat{\rho}_A$ is obtained by tracing over the complement of $A$. 

Note that the estimate in Eq. (\ref{eq:parities}) has significant statistical errors \cite{Unmuth-Yockey:2016znu} and is sensitive to readout errors as a single flip changes the parity. 
 In addition, the practical preparation of the twin copies and interfering them is complicated and challenging to scale.

In view of these difficulties, it has been attempted to estimate the quantum entanglement using the easily available bitstrings, sequence of 0 and 1 from measurements, from a single copy of the system. This motivates our diagnostic strategy: rather than directly estimating entanglement on hardware, we evaluate which features of the measured bitstring distribution are stable under realistic device noise and finite-shot sampling. 
According to Holevo’s inequality, the mutual information of the bitstrings provides a lower bound on the quantum von Neumann entanglement \cite{Holevo:1973,nandc,jpch10}.  It has been shown that for a chain of Rydberg atoms this bound is reasonably tight \cite{lower}. Another option is that a randomized protocol can be applied to compute entanglement entropy \cite{brydges2019probing} from a single copy of the system. However, this protocol is also difficult as this requires a large number of sampled quantum circuits at larger Trotter steps \cite{Asaduzzaman2024capturing} and requires creating global random unitaries which are resource intensive.

The bitstrings from the Rydberg arrays can be obtained \cite{ymzenodo,previouszenodo,currentzenodo} from publicly available facilities \cite{wurtz2023aquila}. Furthermore, it was observed \cite{Kaufman:2024xnu} that a filtering procedure, namely removing the bitstrings with a probability lower than some $p_{min}$ and recalculating the mutual information for the renormalized subset, can lead to closer estimates in the case of a ladder of Rydberg atoms. In a significant majority of cases considered, choosing $p_{min}$ near the inflection point of conditional entropy for the bitstring led to improved agreement with the von Neumann entanglement entropy. We use this filtered mutual information primarily as a practical diagnostic for device data.
In this framing, agreement with ideal-state entanglement provides a useful diagnostic indicator, while the primary output is a quantitative assessment of where Aquila performance departs from theory and which error channels dominate those departures. 

In this article, we discuss the errors associated with the bitstring measurements obtained with the Aquila device and how they affect downstream information measures extracted from those bitstrings, including the filtered mutual information used here as a diagnostic. Our study is focused on the ladder model with 6, 8 and 10 rungs. We analyze sets of 1000 shots.

The article is organized as follows. 
The basic model, a ladder of Rydberg atoms, is introduced in Sec. \ref{sec:ladder}. Bitstrings and mutual information are reviewed in Sec. \ref{sec:MI}.
In Sec. \ref{sec:cumul}, we introduce the cumulative probability distribution as a way to compare bitstring distributions and understand the role of low probabilities. 
The volume effects are discussed in Sec. \ref{sec:vol}. We show examples where the largest value of the probabilities decays exponentially with the size of the system.
Four known sources of errors with Aquila are discussed in \ref{sec:errors}.
Results for Aquila and DMRG comparisons, including filtered mutual-information diagnostics, are presented in Sec. \ref{sec:vne}. Appendices ~\ref{app:dos}-~\ref{app:longer_schedule} provide more information about tests of readout correction methods and alternate state preparation. Additional multipartite analyses are provided in Appendix~\ref{app:multipartite}.

\section{Rydberg simulators and lattice gauge theory}
\label{sec:ladder}

 The specific example discussed here is a ladder of Rydberg atoms. 
Several proposals map lattice gauge models onto ladder geometries in atomic platforms \cite{Bazavov:2015kka,Zhang:2018ufj,Meurice:2021pvj}. We follow the two-leg ladder construction introduced in \cite{Meurice:2021pvj}, which encodes electric-field quantum numbers on the rungs using a spin-1 truncation. While this ladder is not an exact realization of scalar QED \cite{Zhang:2023agx}, it has a rich phase diagram that includes incommensurate (floating) phases \cite{floating}. This makes it a useful testbed that can be readily implemented and simulated on current hardware, enabling benchmarking of Aquila and analysis of device limitations. 

The Hamiltonian of a Rydberg atom array has the general form
\beq
\label{eq:genryd}
H = \frac{\Omega}{2}\sum_i(\ket{g_i}\bra{r_i} + \ket{r_i}\bra{g_i})-\Delta\sum_i  n_i +\sum_{i<j}V_{ij}n_in_j,
\enq
with van der Waals interactions $V_{ij}=\Omega R_b^6/r_{ij}^6$,
given a distance $r_{ij}$ between the atoms labeled as $i$ and $j$. 
We use $\hbar = 1$ units. We use the convention that $r_{ij}=R_b$, the Rydberg blockade radius,  when $V_{ij}=\Omega$, the Rabi frequency. We define the Rydberg occupation  with $n_i\ket{r_i}=\ket{r_i}$ and $n_i\ket{g_i}=0$. 
In the following, we focus on the case of a two-leg ladder. 
The lattice spacing between the rungs is denoted as $a$. 
The value of the rung length (with two atoms each) in lattice spacing units is denoted $\rho$. Here, we focus on the case $\rho=2$ as in Ref. \cite{floating}. 
\def\nr{N_{r}}
We call the number of rungs of the ladder $\nr$.

We use positive detuning which favors the appearance of the Rydberg state. Except for heatmaps where we scan values of $\Delta$, we picked $\Delta/\Omega=3.5$ as in the experimental study of Ref. \cite{floating}. We also often use the value $R_b/a=2.35$ because of the rich near-critical behavior found near this value \cite{floating}. 
For remote simulations of Rubidium atoms using Aquila, we used the value $a=4.1\mu$m because smaller values can be problematic and adjusted $\Omega$ in order to get the desired $R_b/a$. For instance for $R_b/a=2.35$, $\Omega \simeq 6.775$ MHz. The raw data from Aquila in Refs. \cite{lower} and \cite{Kaufman:2024xnu} is available in Refs. \cite{ymzenodo} and \cite{previouszenodo}, respectively, and in Ref. \cite{currentzenodo} for the present work.

\section{Mutual information from a single copy}
\label{sec:MI}
\subsection{Bitstrings and mutual information}

Our starting point is the ground state of the ladder Hamiltonian: $\rho_{AB}=\ket{vac.}\bra{vac.}$. 
Unless otherwise specified, we consider an even number of rungs $\nr$ and the $A,B$  partition splits these rungs into two equal sets.
If the long direction of the ladder is displayed horizontally, $A$ and $B$ will be seen as the left and right halves of the ladder. 
The reduced density matrix is obtained as $\rho_A={\rm Tr}_B\rho_{AB}$ and the 
von Neumann entanglement entropy
\beq
S_{A}^{vN}\equiv-{\rm Tr}(\rho_A \ln(\rho_A)).
\enq
Quantum states will be represented in the Rydberg state occupation basis on each site, which can be represented as a bitstring where $\ket{g} \rightarrow 0, \ket{r} \rightarrow 1$. The probability $p_{\{n\}}$ for the $\{ n \}$ bitstring is the diagonal element of the density matrix
$\bra{\{n\}}\rho_{AB}\ket{\{n\}}$. 
Factoring the state across the $A,B$ partition also divides the bitstring $\{ n \} = \{n\}_A \{n\}_B$, The 
reduced bitstring probabilities, also called marginal probabilities, are obtained from the diagonal elements of the reduced density matrix
\beq
\bra{\{n\}_A}\rho_{A}\ket{\{n\}_A}=p_{\{n\}_A}=\sum_{\{n\}_B}p_{\{n\}_A \{n\}_B}.
\enq
The quantum devices under consideration will yield samples from these bitstring probability distributions, and the structure of these samples will contain some information about the structure of the quantum state. 

The bitstring probabilities can be the subject of classical information quantities such as the total Shannon  entropy,  
\beq
S_{AB}^{ X}\equiv -\sum_{\{n\}} p_{\{n\}}\ln(p_{\{n\}}),\enq
the entropy of the marginals, 
\beq
S_{A}^{X}\equiv -\sum_{\{n\}_A} p_{\{n\}_A}\ln(p_{\{n\}_A}), 
\enq
and the mutual information $I_{AB}^X$ which is non-negative and a lower bound for the entanglement entropy \cite{Holevo:1973,nandc, jpch10}
\beq
\label{eq:mi}
0\leq I^X_{AB}\equiv S_A^X+S_B^X-S_{AB}^X\leq S_A^{vN}.
\enq

Recently, it has been found empirically that this lower bound is reasonably sharp for chains and ladders of Rydberg atoms. \cite{lower, Kaufman:2024xnu}

\begin{figure}[htbp]
\centering
\includegraphics[width=0.7\linewidth]{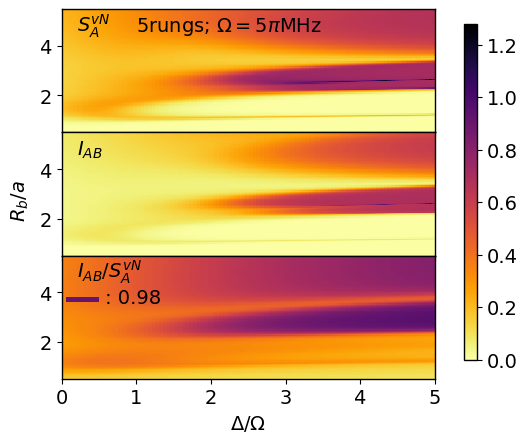} 
\caption{\label{fig:tight} $S^{vN}_A$, $I_{AB}^X$ and   $I_{AB}^X/S^{vN}$ for a 5-rung ladder}
\end{figure}

Fig. \ref{fig:tight} shows this behavior for a 5-rung ladder. The top panel plots the von Neumann entanglement entropy ($S_A^{vN}$), and the middle panel plots the mutual information ($I_{AB}$). They share the same overall pattern across parameter space, including the same high-correlation regions. That agreement suggests ($I_{AB}$) captures the main structure of the entanglement between (A) and the rest of the system.
The bottom panel shows the ratio ($I_{AB}/S_A^{vN}$). This ratio stays near 1 in the regions where the top panel shows substantial entanglement entropy. In those regions, ($I_{AB}$) nearly saturates the bound and gives a quantitatively tight estimate of the entanglement.

As the entanglement entropy goes to zero, the ratio drops and can become much smaller than 1. That is expected: ($I_{AB}$) is a lower bound, so it vanishes at least as quickly, and typically more quickly, than ($S_A^{vN}$). In these same regions, though, the entanglement is already near zero, so even a small ratio still corresponds to a good indication of a vanishing quantity. In that sense, the bound remains effectively tight across the phase diagram, and ratio-level deviations only appear where entanglement is negligible.

Our use of mutual information here is motivated by practicality and diagnostic power. It is directly computable from the same bitstring histograms produced by current hardware, avoiding full-state tomography, twin-copy interference protocols, and large randomized-circuit overheads. 
At the same time, the identical quantity can be evaluated from exact diagonalization and DMRG distributions, giving a controlled comparison between hardware and theory. 
Because $I_{AB}^X$ depends on both marginal and joint probabilities, it is sensitive to redistribution of probability weight across the measured distribution rather than only to the single most probable bitstrings. 
For finite-shot data, this makes filtered mutual information a low-cost but informative observable for tracking device performance.

Further discussion on the extension of this idea to multipartite configurations is discussed in Appendix ~\ref{app:multipartite}.

\subsection{Entanglement entropy estimate from optimal filtration}
We briefly review the filtration-based estimator because it is used later as a practical diagnostic on device data.
\begin{figure}[htbp]
\centering
\includegraphics[width=0.68\linewidth]{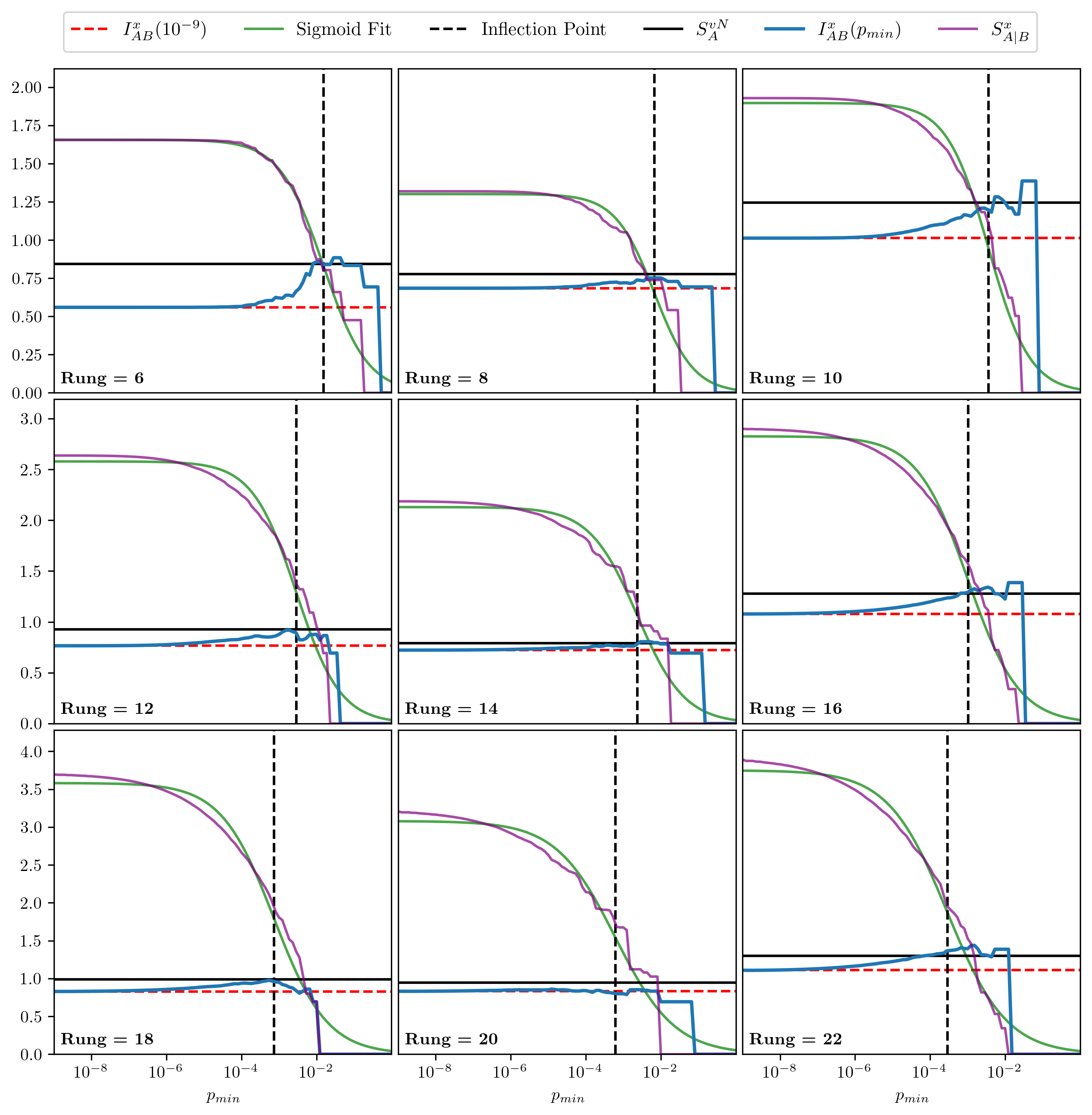} 
\caption{\label{fig:opt} Effects of the number of rungs  on $I_{AB}^X(p_{min})$ for $R_b/a=2.35$, 6,8...22 rungs  obtained via exact diagonalization; comparing sigmoid and mid height approach.}
\end{figure}

Using the discrete probability distribution of bitstrings as a classical summary of the state has clear limitations, because the full quantum information cannot be reproduced by a classical channel. However, in many cases we have assessed, we have found benefit in removing the bitstrings with the fewest counts and normalizing the distribution before the calculation of mutual information. In most cases, this incrementally raises the value of $I_{AB}^X$ closer to $S^{vN}_A$. However, if one removes too many bitstrings, the entire structure of the original state is completely lost, which results in far undershooting or overshooting of $S^{vN}_A$. 

To filter the distribution effectively requires a scheme for determining when to stop. We have found the conditional entropy, another measure of correlation between subsystems, to provide a useful metric for sufficient filtering. The conditional entropy is defined as
\beq S^X_{A\vert B} \equiv S^X_{AB} - S^X_B
\enq
and under filtering, it reduces gradually from an asymptotic initial value to zero. This curve is often well fit with a sigmoid (shifted hyperbolic tangent), the inflection point of which often lands very close to the optimal truncation value. This also closely corresponds to the location where $S^X_{A\vert B}$ has been reduced by half. Fig. \ref{fig:opt} illustrates both indicators with vertical lines for a range of volumes. In practice, the sigmoid fit converges robustly and the inferred inflection point is stable, so $p_{\min}^\star$ is not a freely tunable knob. Importantly, $p_{\min}^\star$ is determined without reference to $S^{vN}_A$ and therefore cannot be chosen post hoc to improve agreement. 
The two exceptions to these methods are when the initial mutual information is very small (indicative of a nearly zero-entropy state) and when $I_{AB}^X$ does not substantially increase under filtering. In both cases, the best available estimate is the initial value of $I_{AB}^X$.
Current quantum devices are limited in their effectiveness by various sources of error including readout errors (the tendency for quantum devices to misread a qubit) and discretization error (approximation of a continuous distribution with samples). Both of these forms of error will be more significant in the bitstrings with fewer counts and will be significantly alleviated by the filtering process. But the demonstrated effectiveness of filtering even on exact probabilities shows that filtering is not only an error-mitigation tool, but also a more general improvement for estimating entanglement entropy. In particular, it is robust to error sources that only modestly perturb the leading probabilities, such as finite-sampling noise and moderate readout errors.

\section{Cumulative probabilities}
\label{sec:cumul}
\subsection{Observations of individual states}

\def\pn{p_{\{ n\} }}
\def\kn{\ket{\{ n\} }}
\def\pla{p_{\Lambda}}
If we focus on the observation of a given state $\kn$ with a probability $\pn$ versus its non-observation with a probability 1-$\pn$, we know from elementary probability that if this Bernoulli process is repeated $\nsh$ times, we expect on average $\nsh \pn$ observations with a standard deviation $\sqrt{\nsh \pn (1-\pn)}$.
It is clear that if $\nsh \pn << 1$, then its physical effect is negligible assuming the state does not carry exceptionally large values of any observable.

From this elementary remark, one might conclude that states with low probabilities can be ignored \textit{en masse}. 
However, if a large group of low-probability states can contribute coherently to an observable, one needs to estimate their number before drawing conclusions. 
A simple example where this point is relevant is the estimation of the entropy. 

\subsection{Cumulative probability distribution}

A simple way to characterize the importance of the low-probability states is to calculate their probability weighted sum below some maximal value $\pla$
\beq
\label{cum_sum}
\Sigma(\pla)=\sum_{\{n\}:\pn \leq \pla} \pn.
\enq
This quantity is the probability to observe any state having a probability less than $\pla$. If we sample the distribution $\nsh$ times, we expect to observe $\nsh \Sigma(\pla)$ states having a probability less than $\pla$. 

This cumulative probability distribution can be estimated by using $\nsh$ measurements and counting the number of times $N_{\{n\}}$ a state $\ket{\{n\}}$ is observed.
We then have an approximation for the probability 
\beq
\pn \simeq  N_{\{n\}}/\nsh.
\label{eq:pest}
\enq

\subsection{Sampling errors}

\begin{figure}[htbp]
\centering
\includegraphics[width=0.65\linewidth]{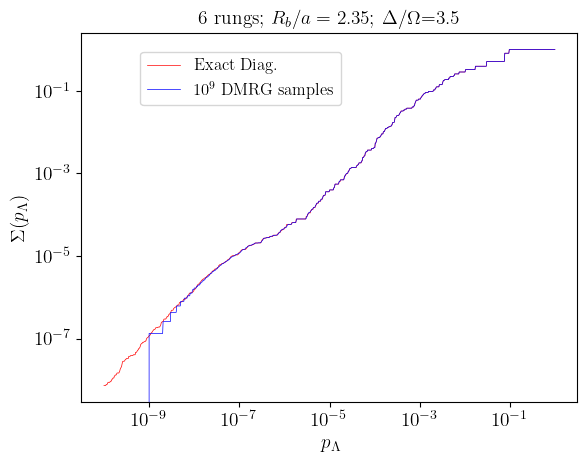} 
\caption{Cumulative probability distribution for 6 rungs $R_b/a=2.35$ and $\Delta/\Omega$=3.5. We used exact diagonalization compared with  $10^9$ DMRG samples.}
\label{fig:6rungsA}
\end{figure}

\begin{figure}[htbp]
\centering 
\includegraphics[width=0.65\linewidth]{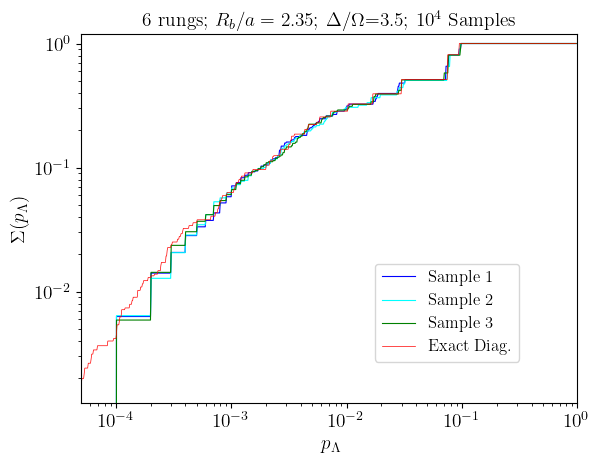} 
\caption{\label{fig:6rungsB}Cumulative probability distribution for 6 rungs $R_b/a=2.35$ and $\Delta/\Omega$=3.5. We used 3 independent DMRG $10^4$ samples obtained by resampling the $10^9$ DMRG samples.}
\end{figure}

If we have a sample with $\nsh$ bitstrings, the probabilities can only be approximated with a resolution $1/\nsh$.
More generally, the relative error on the individual probability estimate Eq. (\ref{eq:pest}) is controlled by $1/\sqrt{N_{\{n\}}}$. This is illustrated  in Fig. \ref{fig:6rungsA} for six rungs where the probabilities for the 4096 states can be calculated very precisely using exact diagonalization. In Fig. \ref{fig:6rungsA}, 
we compare exact diagonalization with $10^9$ density matrix renormalization group (DMRG) samples obtained with ITensor \cite{itensor}. The methods used are explained in Ref. \cite{corona_thesis}. The finite number of samples implies that $10^{-9}$ is the smallest nonzero estimated probability, followed by $2\times 10^{-9}$ and so on. The discrete steps are clearly visible on the lower-left part of the figure and become more difficult to see as we increase $p_{\Lambda}$. For the intermediate and largest values of 
$p_{\Lambda}$, it is clear that DMRG provides accurate approximations. It is also possible to select a subsample statistically distributed according to the original sample. This can be accomplished, for instance, by assigning to each bitstring a subinterval of the $[0,1]$ interval with a length equal to its probability. By using random numbers uniformly distributed over $[0,1]$, we can generate shorter sequences which on average will have the same distribution. 
Three examples of $10^4$ samples obtained with this procedure are shown in Fig. \ref{fig:6rungsB}. Again we see clearly discrete steps corresponding to bitstrings observed once, twice etc. The statistical error on the height can be estimated from the subsample variations.

\newpage

Appendix ~\ref{app:dos} introduces a density-of-probability analysis that extends to larger system sizes. It shows that as volume increases, the cumulative distribution shifts left and becomes more step-like: low-probability states rapidly saturate the cumulative sum, while the maximum probability decreases. This makes exact cumulative-distribution recovery from finite samples increasingly difficult, motivating the volume-scaling analysis in Sec. \ref{sec:vol}.

\FloatBarrier
\section{Volume effects}

\label{sec:vol}
\begin{figure}[htbp]
\centering
\includegraphics[width=\linewidth]{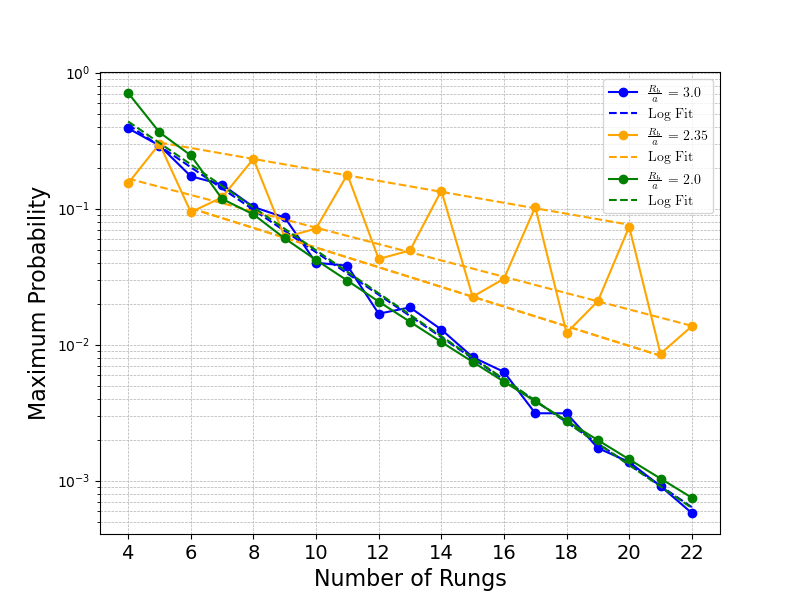}
\caption{Maximum probability versus rung size, with exponential fits of the form $A \cdot e^{-k \nr}$.}
\label{max_prob_fitted}
\end{figure}

We are now in position to study how the previously observed features change when we increase the number of rungs. 
We used DMRG to generate ($10^9$) sampled shots of the Rydberg-array system for ladders with $\nr$ rungs. $\nr$ =22 is the DMRG volume-scaling cutoff used for numerical trend analysis; Aquila hardware runs in this work were practical only up to $\nr$=18, as discussed in Sec. \ref{sec:errors} ~\cite{corona_thesis}. Each raw count was normalized by the total number of shots to yield a probability distribution. 
To describe this effect more precisely, we recorded the maximum probability as a function $\nr$.  We also repeated this calculation for 
$R_b/a$= 2.0 and 3.0.
To characterize how this peak probability decays with system size, we fitted the dependence of the maximum probability on $\nr$ to an exponential form,
\begin{equation}
\label{eq:pmax_decay}
P_{\max}(\nr) = A\,e^{-k\,\nr},
\end{equation}
performing a separate fit for each dataset.  The resulting amplitudes $A$ and decay constants $k$ are summarized in Table \ref{tab:fit_params}, and the curves are overlaid on the data in Fig. \ref{max_prob_fitted}. For further reference, the exponential factor $e^{-k\,\nr}$ can be written as $0.835^{N_{atoms}}$ for $R_b/a=3.0$ and 
$0.955^{N_{atoms}}$ for $R_b/a=2.35$ with $\nr\mod{3}=2$. 

The results of the fit are shown in Table \ref{tab:fit_params}.
\begin{table}[htbp]
  \centering
  \setlength{\tabcolsep}{12pt}  

  \caption{%
    Exponential‐decay fit parameters for the maximum probability,
    $P_{\max}(\nr)=A\, \cdot e^{-k_{e}\,\nr}$, at various blockade‐to‐spacing ratios $R_b/a$.%
  }
  \label{tab:fit_params}
\begin{tabular}{|c|c|c|c|}
    \toprule
    $R_b/a$   &$\nr$         & $A$     & $k_e$       \\
    \midrule
    $2.0$   & All           & $1.87$  & $0.182$   \\
    2.35& $\ (\nr\mod{3})=1$   & $0.289$ & $0.069$   \\
    2.35& $\ (\nr\mod{3})=2$   & $0.487$ & $0.093$ \\
    2.35&$\ (\nr\mod{3})=0$   & $0.273$ & $0.083$   \\
    3.0 &      All       & $1.743$ & $0.180$  \\
    \bottomrule
  \end{tabular}
\end{table}
\FloatBarrier

These results make clear that in general an exponential decay of the largest probability is observed. A faster rate without the modulo 3 effect is observed for $R_b/a$= 2.0 and 3.0. 
We can also fit
\begin{equation}
  P_{\max}(N_r)=A \cdot 2^{-k_{2}N_r}
\end{equation}

\noindent
which directly compares the scaling of $(P_{\max})$ with the Hilbert‑space dimension. From this fit we find ($k_2$=0.262) for ($R_b/a=2.0$) and ($k_2$=0.259) for ($R_b/a=3.0$). For ($R_b/a=2.35$), ($k_2$=0.10) (mod 0), ($k_2$=0.067) (mod 1), and ($k_2$=0.12) (mod 2).

The implication of this exponential decay is important. 
If we have only $\nsh$ shots, then there is a system size such that the largest probability will be $1/\nsh$, which means that any state is observed once or zero times.
This implies large error bars on the mutual information and can only be remedied by using more shots. In summary, the 
behavior observed in this section implies that the cost of measuring the mutual information increases exponentially with the volume.

\section{Sources of Error}

\label{sec:errors}
We will now repeat the calculations of bitstring probabilities using public Rydberg-array facilities. We have used the Aquila device described in the following whitepaper \cite{wurtz2023aquila}. Before presenting the results, we need to address four known sources of errors. These effects appear chronologically during the device operation in the order listed below.  
\begin{enumerate}
\item
Sorting fidelity (missing atoms at the beginning)
\item
Adiabatic preparation (possible level crossings and transitions during the rampup)
\item
$\Omega$ rampdown (should be fast enough as opposed to adiabatic)
\item
Readout errors (there is a 0.01 probability for $g\rightarrow r$,  and 0.08 probability for $r\rightarrow g$ )
\end{enumerate}
\newpage

 We discuss them separately in the subsections below. The executive summary is that: 
 \begin{enumerate}
\item Sorting fidelity: incorrect presequences  can be removed by postselection. This leads to data loss that does not exceed 40\% for the cases considered here.
\item Adiabatic preparation: this is probably the main cause of errors. Longer rampup times seem to improve the situation for numerical calculations.
\item $\Omega$ Rampdown: a 0.05 $\mu$s rampdown is possible and only causes small errors compared  to a slower process.
\item Readout errors: they are very significant but they can be mitigated very efficiently after getting the bitstrings.
\end{enumerate}
\subsection{Sorting fidelity}
When the atoms are loaded and moved to their desired position, it is not uncommon to lose a few of them due to finite trap lifetime, background-gas collisions, motional heating, or Rydberg excitation/decay processes that eject atoms from the tweezer.
Fortunately, incorrect presequences can be identified in the 
data file provided and the corresponding measurements can be discarded. 
The likelihood that a user pattern is successfully generated is called the sorting fidelity which is estimated by the number of correct presequences divided by the number of shots.  
An example is provided for ladders with 6, 8, 10, ..., 18 rungs in Table \ref{tab:states}. We extended Aquila runs up to 18 rungs, which was the largest size
practical at the time.
\begin{table}[htbp]
\centering
 \begin{tabular}{|c|c|c|c|c|c|c|c|}
\hline
Number of atoms&12&16&20&24&28&32&36\\
\hline
Correct presequences&730&763&734&694&686&661&562\\
\hline
\end{tabular}
\caption{\label{tab:states}Number of Correct presequences for 1000 shots with Aquila, in Oct. 2024.}
\end{table}

Apparently, the sorting fidelity  has been improving with time. In October 2024, we obtained 730 correct presequences with 6 rungs and the  parameters used in the previous sections. We ran the same code 5 times around  January 2025 and we obtained  885, 883, 863, 884, and 884, 
correct presequences. 

\begin{figure}[htbp]
\centering
    \includegraphics[width=0.95\linewidth]{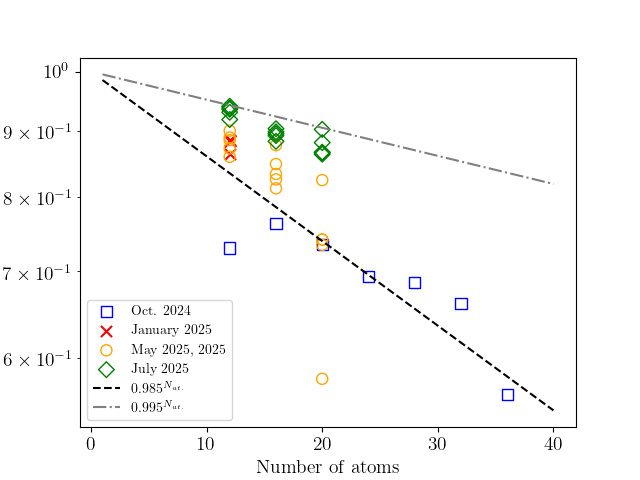}
    \caption{Fraction of correct presequences vs. number of atoms. The straight lines represent $0.995^{N_{atoms}}$ and $0.985^{N_{atoms}}$ which are typical for July 2025 and October 2024 respectively.}
    \label{sortingfidelity}
\end{figure}

The sorting fidelity decreases exponentially with the system size. For instance the October 2024 sorting fidelity can be approximated with $0.985^{N_{atoms}}$. Better sorting fidelities were obtained in July 2025. Several examples are shown in Fig. \ref{sortingfidelity}. Note that the presequences shown here are for the ladder configuration defined in Sec.~\ref{sec:ladder}.

\newpage

In summary, the current sorting fidelity could be as good as 
$0.995^{N_{atoms}}$. For a 20 by 20 square array, it means that in principle we need to discard about 86\% of the data. For linear structures like chains and ladders, a single missing atom partially disconnects its left and right neighbors. However the effects may be less drastic for square arrays. It would be interesting to study the comparison of the features of measurement with correct and incorrect presequences. 

\subsection{Adiabatic ramping up}
\label{sec:adiabatic rampup}
In our study, we prepare the ground state at various parameters by applying quantum adiabatic approximation \cite{born1928beweis,kato1950adiabatic} which states that a system initially prepared in an eigenstate of a Hamiltonian, remains at the same instantaneous eigenstate as we vary the parameters of our Hamiltonian ``slowly''. In our simulation, we prepare the ground state $|0\rangle^{\otimes N}$ at the disordered phase and then slowly vary Rabi frequency and detuning to the target values to prepare the new ground state. The slowness is met by various adiabatic conditions \cite{Albash:2016zte}, where in the most crude version, the total evolution time requires to be greater than the square of the minimal of the inverse gap during adiabatic preparation. When the desired ground state is prepared, one needs to turn off $\Omega$ as fast as possible (this will be discussed in the next subsection) and then measure the occupation of the individual atoms. 

During the adiabatic evolution, we have an instantaneous Hamiltonian 
\beq
\label{eq:instH}
H(t)=H _n(t)+H_{\Omega} (t),
\enq
where the
diagonal part of the Hamiltonian in the computational basis is denoted 
\beq
H _n(t)= -\Delta(t)\sum_i  n_i +\sum_{i<j}V_{ij}n_in_j,
\enq
and the Rabi term
\beq
H_{\Omega} (t)= \frac{\Omega (t)}{2}\sum_i(\ket{g_i}\bra{r_i} + \ket{r_i}\bra{g_i})
\enq
In most of the simulations performed here, we used trapezoidal time profiles for $\Omega (t)$ and the regular Piecewise Linear (PWL) form for $\Delta(t)$ (see Fig. 4 in \cite{lower} for an illustration) with a standard 
$4\,\mu$s schedule given in table \ref{original_time_series}.
    \begin{table}[htbp]
    \centering
    \begin{tabular}{|c|c|c|c|c|}
    \hline
        $t$ & $0$ & $0.5\,\mu\mathrm s$ & $3.95\,\mu\mathrm s$ & $4\,\mu\mathrm s$ \\ \hline
        $\Omega(t)$ & $0$ & $\Omega_{\max}$ & $\Omega_{\max}$ & $0$ \\ \hline
        $\Delta(t)$ & $-\Delta_{\max}$ & $-\Delta_{\max}$ & $+\Delta_{\max}$ & $+\Delta_{\max}$ \\ \hline
    \end{tabular}
    \caption{$4 \mu s$ ramp schedule. }
    \label{original_time_series}
\end{table}
 
The ideal evolution for this time-dependent Hamiltonian can be calculated using the  
time-ordered exponential which can be approximated by a Trotter sequence
\beq
\begin{split}
\ket{\Psi(t)} &= \prod_j \exp\!\left(-iH_{\Omega}(t_j)\delta t\right) \\
&\quad\times \exp\!\left(-iH_n(t_j)\delta t\right)\ket{0,0,\dots 0}
\end{split}
\enq
for equally spaced time steps
$t_j=0, \delta t, 2\delta t, \dots t_{final}$
For small systems (up to 5 or 6 rungs) the calculation of the evolution and the probabilities to be in any of the states of the instantaneous Hamiltonian can be performed with great precision.

In the rest of this subsection we focus on the impact of the adiabatic ramping up. The ramping down of $\Omega$ is discussed in the next subsection.
From empirical observation of numerical calculations, the ramping up of $\Omega$ is uneventful at large negative detuning while more complex features are observed when $\Delta$ is ramped up. 
Significant loss of fidelity to the instantaneous ground state occurs when energy gaps become small, particularly for excited states that couple strongly to the vacuum via the time-dependent perturbation operator $-\Delta(t)\sum_j \hat n_j$.
The problem occurs for some special values of $\Delta$ and sometimes the fidelity can ``recover" for larger times of the rampup. The issue can be mitigated by using a slower ramping up in the difficult region.

\newpage

\begin{figure}[htbp]
\centering
    \includegraphics[width=0.66\linewidth, trim= 0 0 14 0, clip]{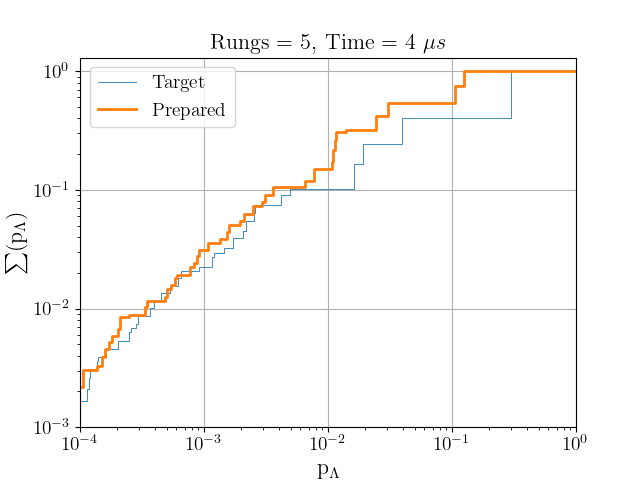}  
    \caption{\label{fig:4vstarget}Comparison of Cumulative distribution for the adiabatically prepared state with a 4$\mu$s time pulse compared to target (5 rungs). }
\end{figure}

\begin{figure}[htbp]
\centering
    \includegraphics[width=0.66\linewidth, trim= 0 0 14 0, clip]{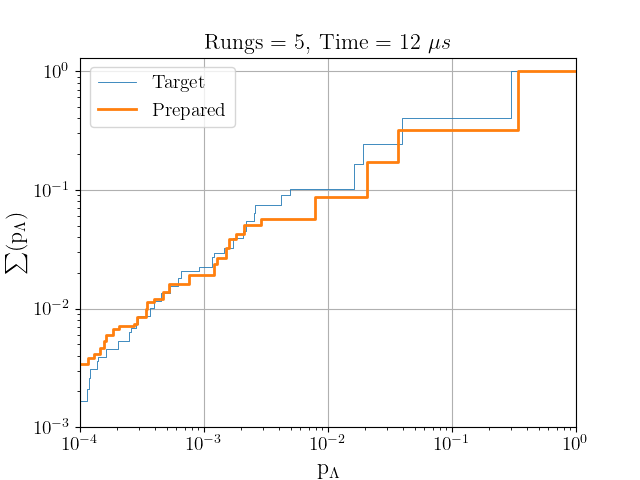}  
    \caption{\label{fig:12vstarget}Comparison of Cumulative distribution for the adiabatically prepared state with a 12 $\mu$s time pulses compared to target (5 rungs). }
\end{figure}

As an illustration, we considered the standard example $\rho=2.0$, $R_b/a$=2.35, $a=4.1\mu m$ and $\Delta=3.5\Omega$. This corresponds to $\Omega=1.078224\ \times 2\pi$ MHz=6774683 Hz.
We used $\delta t$=0.02 $\mu$s. The results are shown in Figs. \ref{fig:4vstarget} \& \ref{fig:12vstarget}. Significant departures from the target probabilities are clearly visible for the $4\,\mu$s sequence. However, extending the rampup to $12\,\mu$s goes a long way to fix the problem. We have tested a modified ramp up schedule within the $4\, \mu s$ window in Appendix \ref{app:ramp_variation}. We observe minimal difference between these two schedules. However, increasing the original schedule to $12 \mu s$ does have a significant impact. Figs. \ref{fig:4vstarget} \& \ref{fig:12vstarget} show the $12\,\mu s$ schedule is closer to the target state than the $4\, \mu s$ schedule.

\subsection{Rampdown}
\begin{figure}[htbp]
\centering
    \includegraphics[width=0.7\linewidth, trim=0 0 0 14, clip]{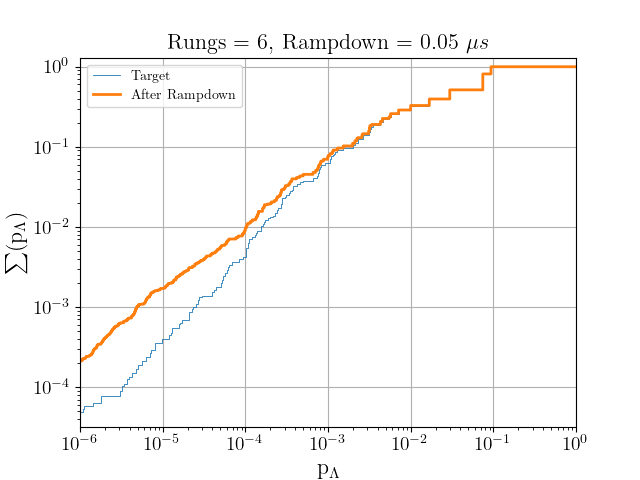}
    \caption{Cumulative probability distribution for the ideal vacuum and its ramped down version for 0.05 $\mu$s.}
    \label{rampdownvariationA}
\end{figure}

\begin{figure}[htbp]
\centering
    \includegraphics[width=0.7\linewidth, trim=0 0 0 14, clip]{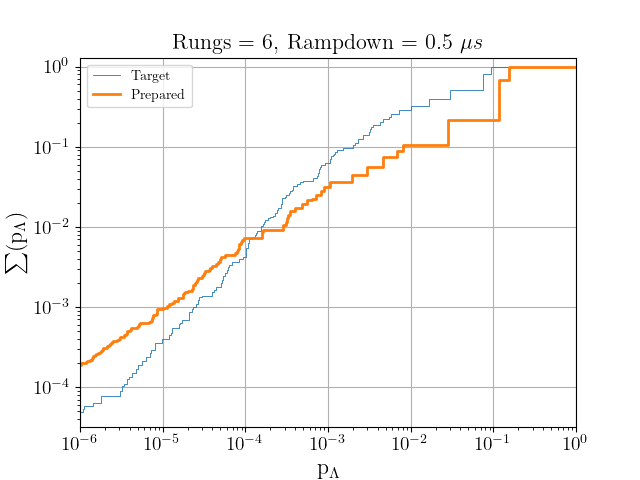}
    \caption{Cumulative probability distribution for the ideal vacuum and its ramped down version for 0.5 $\mu$s.}
    \label{rampdownvariationB}
\end{figure}

An important aspect of the measurement process is switching off Rabi frequency $\Omega$ before measurement. Ideally the rampdown of the Rabi frequency needs to be instantaneous so that the process can be described by a fully diabatic evolution and thus does not alter the ground state at target Rabi frequency. For the range of frequencies we investigated in our study, it has been observed that a ramping down time of $0.5\,\mu$s may be too slow \cite{lower}, whereas $0.05\,\mu$s should be short enough to faithfully recover the target ground state. The cumulative distribution plots in Figs. \ref{rampdownvariationA} \& \ref{rampdownvariationB} demonstrate the importance of choosing a fast ramping down of the Rabi frequency for the 6-rung ladder.

\FloatBarrier

\subsection{Mitigation of readout errors}
\label{mitigation_of_readout_error}
Readout errors are common in quantum computing devices. According to the Aquila whitepaper \cite{wurtz2023aquila} there is a 0.01 probability for reading $\ket{g}$ as $\ket{r}$  and a 0.08 probability for reading $\ket{r}$ as $\ket{g}$. We believe that these numbers have small temporal variations that can in principle be monitored, however we will not take these temporal effects into account in the following. 

The 0.08 probability for misidentifying the Rydberg state is quite large and for states with large Rydberg occupation in large systems, the chances to misidentify the bitstring can be significant. Fortunately, it is possible to invert the confusion matrix which expresses the observed probabilities as linear combination of the true probabilities. This can be done exactly for small systems and by using the M\textsubscript{3} readout‐error mitigation with error rates $p_{0\to1}=0.01$, $p_{1\to0}=0.08$, as introduced by Nation \emph{et al.} \cite{nation2021scalable}, which builds a reduced assignment matrix over only the observed bitstrings and then performs a matrix-free iterative inversion to recover the unbiased probabilities. 
For an independent readout correction implementation see Ref. \cite{PhysRevA.106.012423}.

Before applying the M\textsubscript{3} code to Aquila data, we have tested its validity by using ideal 
DMRG samples with the same numbers of samples as the Aquila data discussed after and applying bit flips with the known probabilities specified above on each bitstring. We then applied the M\textsubscript{3} method and compared with the original samples. In Appendix \ref{app:dmrg_validation} we show that the agreement for the largest probabilities is quite good.
Importantly, we noticed applying the M\textsubscript{3} correction to a dataset that originally has $N$ unique (distinct) bitstrings will still yield exactly those $N$ unique bitstrings. The mitigation never introduces new strings or drops existing ones.  However, because M\textsubscript{3} is carried out by inverting the readout–error confusion matrix, the corrected probabilities need not respect the “one–count–floor” $1/\nsh$, in other words they are not necessarily integer multiples of $1/\nsh$.  In practice you will often see $p_i^{(\rm mitig)}<1/\nsh$ , which at first glance seems unphysical, but in fact simply reflects how the linear inversion redistributes weight to remove systematic bias rather than a literal extra “fractional” measurement.

\begin{figure}[htbp]
\centering
\includegraphics[width=0.9\linewidth]{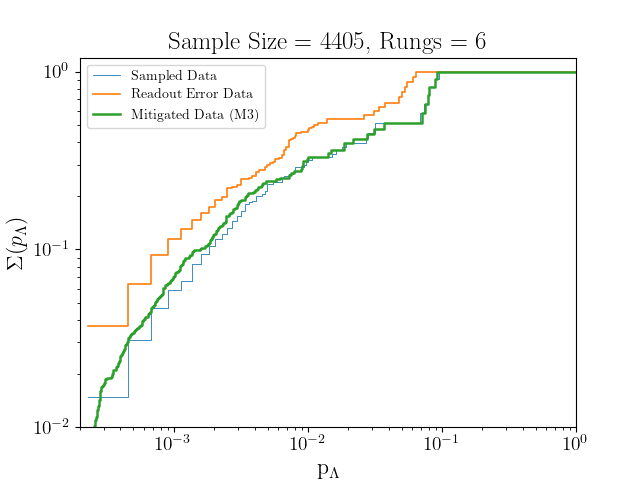}
    \label{fig:6rung-cdf-dmrgA}
    \caption{Comparison of 4405 bitstrings for $R_b/a=2.35$ obtained with DMRG, the readout error version and the M\textsubscript{3} corrected version.}
\end{figure}    

\FloatBarrier
\section{Estimation of the von Neumann entropy}
\label{sec:vne}

In this section, we use the filtered mutual information as a practical diagnostic, as it provides an accessible measure of the entanglement entropy with current limitations, on device data. For each ladder size $\nr\!=\!6,8,10$, we show the cumulative distribution and mutual information side by side for Aquila, alongside DMRG references. In Appendix \ref{app:dmrg_validation} we reproduce the same analyses using only DMRG data with simulated readout noise and mitigation, to validate our M\textsubscript{3} implementation independently of the hardware. We now proceed to discuss the different sizes and their general features.
We restrict this section to 6, 8, and 10 rungs because hardware time and computational resources were limited, and instability already appears at these volumes, making them sufficient to diagnose the dominant error channels.

\subsection{6-, 8-, and 10-Rung Ladders}
\begin{figure}[htbp]
\centering
    \includegraphics[width=0.68\linewidth, trim=0 0 0 14, clip]{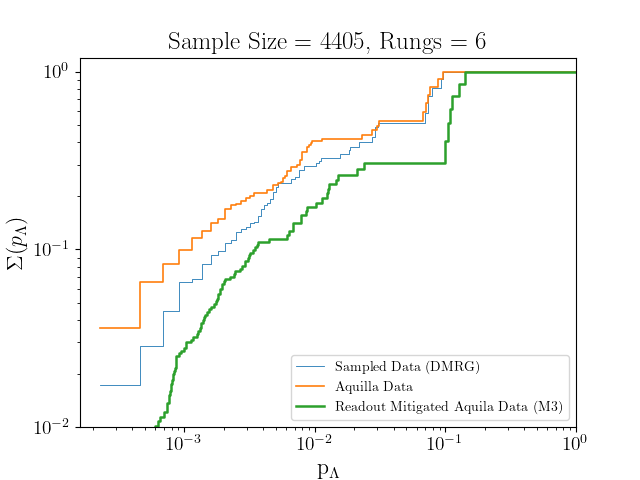}
    \caption{Comparison of cumulative distribution for the 6-rung ladder}
    \label{fig:6rung-cdf}
\end{figure}

\begin{figure}
    \centering
    \includegraphics[width=0.68\linewidth, trim=0 0 0 14, clip]{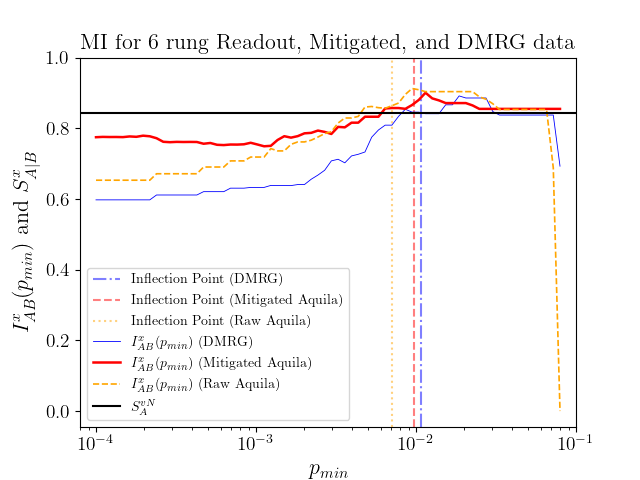}
    \caption{Comparison of mutual information for the 6-rung ladder.}
    \label{fig:6rung-mi}
\end{figure}

As shown in Figs.~\ref{fig:6rung-cdf} and ~\ref{fig:6rung-mi} the raw Aquila cumulative distribution for $\nr=6$ closely matches both the down‐sampled (same size as Aquila) and ideal DMRG ($10^9$ samples) curves at high $p$, and the corresponding MI remains nearly on top of the DMRG result across all truncation thresholds.  Remarkably, the sigmoid inferred truncation point for the raw, mitigated, and DMRG curves all coincide with the exact von Neumann entropy, demonstrating that, for this small system, the device is accurately capturing the dominant states.  However, applying M\textsubscript{3} here actually worsens the cumulative distribution: the mitigated curve deviates further from the ideal DMRG than the unmitigated data.

\begin{table}[htbp]
  \centering
  \begin{tabular}{|c|c|c|}
    \toprule
    Method & $p_{\min}^\star$ & $I_{AB}(p_{\min}^\star)$ \\
    \midrule
    Exact von Neumann entropy        & —                  & 0.8441 \\
    DMRG inflection point       & $1.09(6)\times10^{-2}$ & 0.85(1) \\
    Mitigated Aquila inflection      & $9.71\times10^{-3}$ & 0.87   \\
    Raw Aquila inflection            & $7.08\times10^{-3}$ & 0.86   \\
    \bottomrule
  \end{tabular}
  \caption{Inflection thresholds and corresponding mutual‐information values for the 6‐rung ladder under raw and mitigated Aquila and DMRG.}
  \label{tab:inflection_6rungs_updated}
\end{table}

\FloatBarrier

\begin{figure}[htbp]
\centering
    \includegraphics[width=\linewidth, trim= 0 0 0 14, clip]{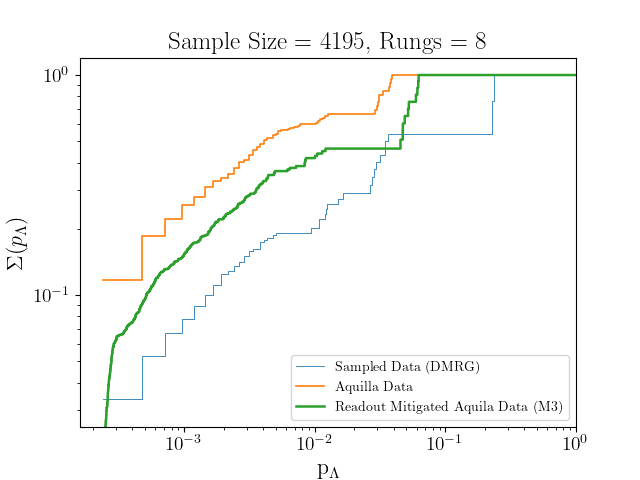}
    \caption{Comparison of cumulative distributions for the 8-rung ladder.}
    \label{fig:8rung-A}
\end{figure}

\begin{figure}[htbp]
\centering
    \includegraphics[width=\linewidth, trim= 0 0 0 14, clip]{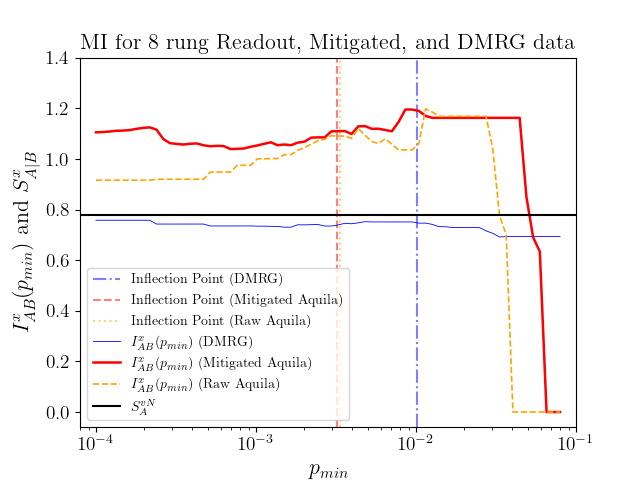}
    \caption{Comparison of mutual information for the 8-rung ladder.}
    \label{fig:8rung-B}
\end{figure}

For $\nr=8$ (Figs.~\ref{fig:8rung-A} \& ~\ref{fig:8rung-B}), the raw Aquila cumulative distribution shows significant deviations, particularly in the highest‐probability region, and although mitigation brings the bulk of the curve closer to DMRG, it fails to recover the true leading probabilities. Correspondingly, the mutual information in Fig.~\ref{fig:8rung-B} overshoots the exact entropy by a wide margin, while the DMRG‐based mutual information remains just below the exact value for all $p_{\min}$.  This indicates that, at this size, readout correction alone cannot compensate for the hardware’s loss of fidelity in the most significant bitstrings.

\begin{table}[htbp]
  \centering
  \begin{tabular}{|c|c|c|}
    \toprule
    Method & $p_{\min}^\star$ & $I_{AB}(p_{\min}^\star)$ \\
    \midrule
    Exact von Neumann entropy        & —                  & 0.7769 \\
    DMRG inflection point            & $1.01(8)\times10^{-2}$ & 0.749(4)   \\
    Mitigated Aquila inflection      & $3.19\times10^{-3}$ & 1.11   \\
    Raw Aquila inflection            & $3.34\times10^{-3}$ & 1.09   \\
    \bottomrule
  \end{tabular}
  \caption{Inflection thresholds and corresponding mutual‐information values for the 8‐rung ladder under raw and mitigated Aquila and DMRG.}
  \label{tab:inflection_8rungs_updated}
\end{table}

\FloatBarrier

\begin{figure}[htbp]
\centering
    \includegraphics[width=0.9\linewidth, trim= 0 0 0 14, clip]{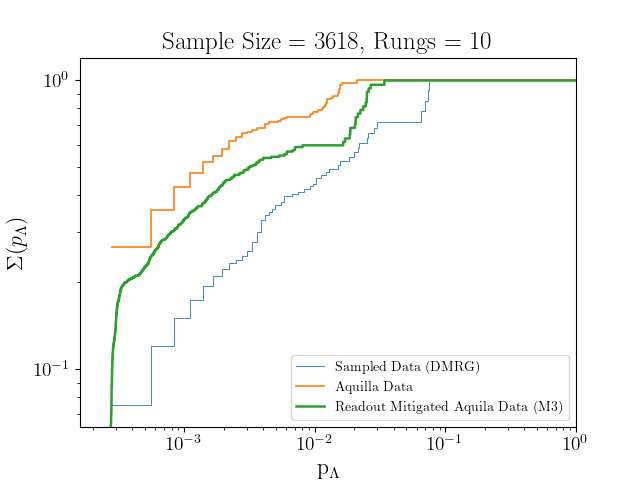}
    \includegraphics[width=0.9\linewidth, trim= 0 0 0 14, clip]{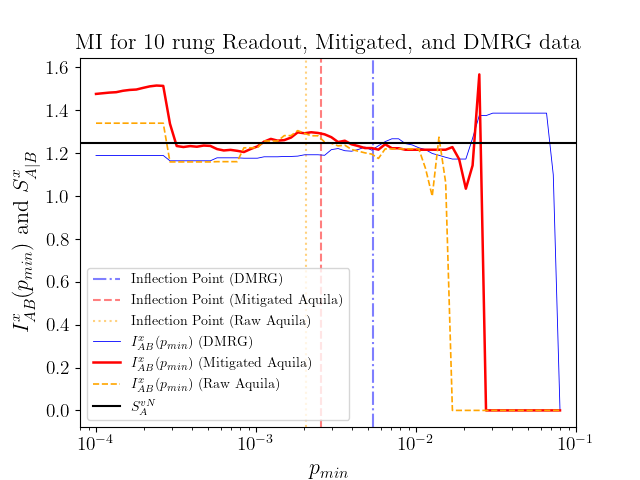}
  \caption{Comparison of cumulative distributions (top) and mutual information (bottom) for the 10-rung ladder.}
  \label{fig:10rung-pair}
\end{figure}

At $\nr=10$ (Fig.~\ref{fig:10rung-pair}), the raw cumulative distribution again departs substantially from DMRG, but here mitigation yields a modest improvement in the tail region.  Interestingly, the mutual information‐based truncation for both raw and mitigated Aquila data still lands close to the exact entropy, even though the overall mutual information curve shapes differ markedly from the DMRG reference.  This suggests that, despite distorted probability profiles, the global mutual information estimator retains some resilience in locating the correct entropy for larger ladders.

\begin{table}[htbp]
  \centering
  \begin{tabular}{|c|c|c|}
    \toprule
    Method & $p_{\min}^\star$ & $I_{AB}(p_{\min}^\star)$ \\
    \midrule
    Exact von Neumann entropy        & —                  & 1.2455 \\
    DMRG inflection point       & $5.01(36)\times10^{-3}$ & 1.26(2) \\
    Mitigated Aquila inflection      & $2.55\times10^{-3}$ & 1.29   \\
    Raw Aquila inflection            & $2.05\times10^{-3}$ & 1.29   \\
    \bottomrule
  \end{tabular}
  \caption{Inflection thresholds and corresponding mutual‐information values for the 10‐rung ladder under raw and mitigated Aquila and DMRG.}
  \label{tab:inflection_10rungs_updated}
\end{table}
\FloatBarrier
\subsection{Validation and Interpretation}

In Appendix~\ref{app:dmrg_validation}, we repeat these exact same plots using only DMRG data with simulated readout noise and M\textsubscript{3} mitigation.  There, the mitigated cumulative distributions align almost perfectly with the ideal curves, and all mutual information traces (noisy and corrected) collapse onto one another with truncation points matching the exact values.  That clean benchmark confirms our M\textsubscript{3} implementation is sound.

Because readout error correction fails to restore the true probability distribution in Aquila, sometimes even making it worse, these results strongly suggest that readout errors are \emph{not} the dominant source of discrepancy.  Our group is therefore investigating other error channels  which may play a larger role in the observed deviations. See Appendices~\ref{app:adiabatic}, \ref{app:ramp_variation}, and \ref{app:longer_schedule} for further discussion of adiabatic‐ramp effects, schedule variations, and extended ramp‐time tests.

The tables of this section make clear that the readout mitigation has a very small effect on the optimally filtered values of the mutual information. This can be partially explained in a simplified readout error model where the bitstrings with large probabilities are mostly depleted by the readout errors and that their enhancement from bitstrings with lower probabilities are negligible. This leads to a simple estimate
\beq
\label{eq:correct}
N_{\{n\}}^{observed}\simeq N_{\{n\}}^{true}(1- p_{1\rightarrow 0})^{n_R} (1-p_{0\rightarrow 1})^{N_q-n_R}.
\enq
We call $(1- p_{1\rightarrow 0})^{n_R} (1-p_{0\rightarrow 1})^{N_q-n_R}$ the ``depletion factor", where $N_q$ is the number of qubits.
The bitstrings with large probabilities tend to have approximately the same $n_R$ and so the depletion factor  is almost the same for these types of states. As an example, for 6 rungs, the 6 largest probabilities have $n_R=4$ 
and a depletion factor 0.66, while the next set of four has $n_R=3$ and a depletion factor 0.71. This can be mitigated by dividing by the depletion factor. In other words the readout errors stretch the cumulative distribution towards the left and shrinks the vertical changes by about the same factor.

This approximate scaling relation appears 
in Figs. \ref{fig:6rung-cdf}-\ref{fig:10rung-pair} where the states with the largest probabilities can be individually paired after the readout correction. As the factor is approximately the same for all large enough probabilities, 
and since the filtering requires to renormalize the probabilities, we see that a common factor for all the high probabilities is absorbed in the renormalization and plays no role in this limit. 

\FloatBarrier
\section{Conclusions}

In summary, we compared Aquila bitstring probability distributions for two-leg ladder gauge-model instances against high-accuracy DMRG and exact numerical references, and identified several sources of disagreement. We find that errors in the cumulative distributions and related information measures arise from finite sampling, sorting fidelity, adiabatic state preparation, ramp-down of the Rabi frequency before measurement, and readout effects.

Except for the state preparation, it is possible to reduce the known source of errors to an acceptable level using simple mitigation techniques as long as the system size is not too large. Until very recently, the adiabatic preparation had to take place within 4$\mu$s. Using numerical integration of the evolution of a time-dependent rampup, we found that increasing the time of preparation could substantially improve the fidelity to the target ground state. 
Very recently, we were able to increase the Aquila preparation time up to 20$\mu$s with the caveat that coherence could suffer and we were not able to reproduce the improvement obtained by numerical methods. Uncertainties in detuning, Rabi frequency, and the atom positions could be related to this observation.
Slower ramp-ups near level-crossings or other mitigation methods \cite{lukin2024,Li:2024lrl} are planned. 

The large-volume behavior of the cumulative probability distribution offers interesting features. The largest value of the bitstring probability appears to decay exponentially with the size of the system. In order to estimate the mutual information with reasonable errors, it is necessary to observe bitstrings with the largest probabilities more than once. This represents a cost that grows exponentially with the size of the system. Similar conclusions can be drawn for the sorting fidelity. 

Additional multipartite analyses based on the weak monotonicity combination in Eq. (\ref{eq:weak}) are provided in Appendix~\ref{app:multipartite}.

\section*{Data Availability}
All raw and processed datasets generated or analyzed in this study are publicly available on Zenodo:  
Aquila‐device measurements \cite{currentzenodo},  \cite{previouszenodo}, and DMRG benchmark samples \cite{corona_data}.

\section*{Acknowledgements} This research was supported in part by the Dept. of Energy
under Award Number DE-SC0019139 and DE-SC0025430.
This research used resources of the National Energy Research Scientific Computing Center (NERSC), a Department of Energy Office of Science User Facility under Contract No. DE-AC02-05CH11231 using NERSC award DDR-ERCAP 0023235. 
We thank K. Klymko, J. Balewski, D. Camps, S. Darbha, E. Rrapaj, M. Kornjaca and P. Lopes 
for their help during the NERSC/QCAN office hours.
We thank 
the Amazon Web Services and S. Hassinger for facilitating remote
access to QuEra through the Amazon Braket while teaching quantum mechanics.
We thank Mao Lin and P. Komar for detailed discussion about the Amazon braket operation of Aquila.
We thank S. Cantu, Sheng-Tao Wang, A. Bylinskii, S. Everett,  and P. Lopes, for technical discussions related to Aquila.
We thank the Department  of Physics and Astronomy and a grant ``Investment in Strategic Priorities" from the Provost Office at U. Iowa for supporting part of the cost of the analog simulations presented here.
We thank the University of Iowa for providing access to the Argon computing facilities for the DMRG calculations. 
This work benefited from the Co-design for Fundamental Physics in the Fault-Tolerant Era  IQus workshop (April 2025), which was supported by U.S. Department of Energy, Office of Science, Office of Nuclear Physics, InQubator for Quantum Simulation (IQuS) under Award Number DOE (NP) Award DE-SC0020970 via the program on Quantum Horizons: QIS Research and Innovation for Nuclear Science. M.A. would also like to thank S. Hassinger, T. Aunon, N. Owens and others on the on-boarding team for arranging quantum credits for the Aquila device through amazon braket.

\newpage

\appendix

\section{Density-of-States View}
\label{app:dos}
\begin{figure}[htbp]
\centering
    \includegraphics[width=0.9\linewidth, trim= 0 0 0 14, clip]{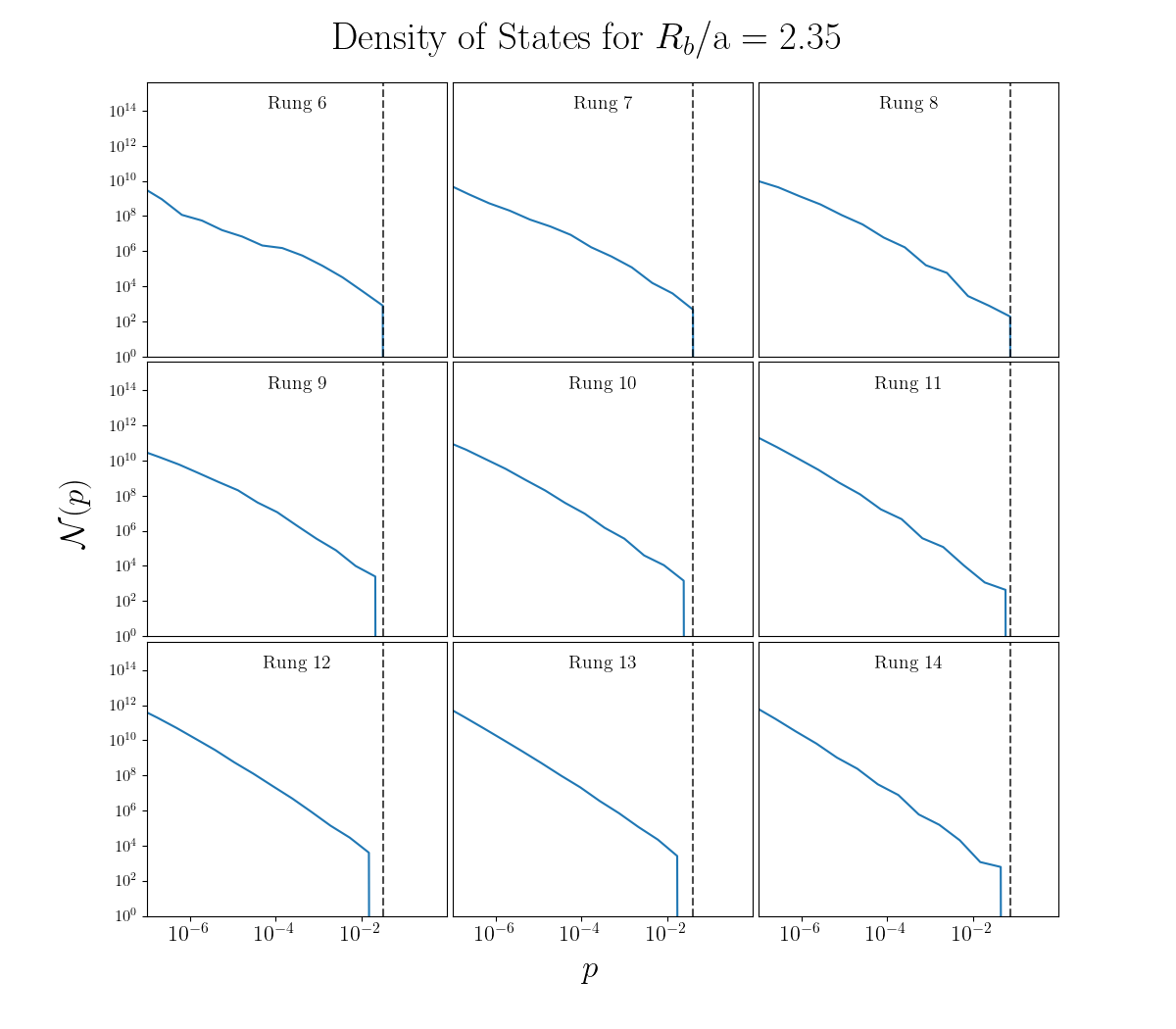}
    \caption{Density of states for $R_b/a = 2.35$, $\Delta/\Omega=3.5$ and number of rungs 6, 8, ...,14.}
    \label{fig:dos_rba_235}
\end{figure}

The density of states for $\nr$ up to 14 is shown in Fig.~\ref{fig:dos_rba_235}. Rather than fixing the bin edges a priori, the discretization of the probability axis is constructed directly from the empirical distribution. Starting from the normalized bitstring probabilities $\{p_i\}$, we first define the cumulative distribution function as in Eq. ~\ref{cum_sum} from the main text.

This cumulative distribution is sampled using an initial coarse binning, after which redundant points where $\Sigma(p)$ has saturated are removed. The remaining points define a set of distinct probability values $\{p_k\}$ that span the support of the data. The resulting number of bins is therefore data dependent and is typically close to 17 for the systems considered here.

Using the minimum and maximum probabilities present in this set, $p_{\min}$ and $p_{\max}$, we define bin centers that are equally spaced in $\log_{10} p$,
\begin{equation}
\label{eq:dos_pk}
p_k = 10^{\,\log_{10} p_{\min} + k\,\Delta(\log_{10} p)}, \qquad
k = 0,1,\dots,N_{\mathrm{bins}}-1,
\end{equation}
with
\begin{equation}
\label{eq:dos_delta_log}
\Delta(\log_{10} p) =
\frac{\log_{10} p_{\max} - \log_{10} p_{\min}}{N_{\mathrm{bins}}-1}.
\end{equation}
To estimate the density of states $\mathcal{N}(p)$ at each bin center $p_k$, we count the number of states whose probabilities lie within a small linear window around $p_k$. Using a half width $\delta p$, this estimator may be written as
\begin{equation}
\label{eq:dos_density_est}
\mathcal{N}(p_k)\,\Delta p
\;\approx\;
\Sigma(p_k + \delta p) - \Sigma(p_k - \delta p),
\end{equation}
with $\Delta p = 2\delta p$ and $\delta p = 0.1$ in the data shown. This procedure yields a numerical estimate of the density of probability states that is well resolved over many decades in $p$ while avoiding empty bins in regions with no support.

\newpage

We observe that the approximate linear behavior persists. Comparing distributions for successive $\nr$ (in increments of 3) shows a slight shift towards the left as the number of rungs increases.

\subsection{Density of probability}
\begin{figure}[htbp]
\centering
\includegraphics[width=\linewidth]{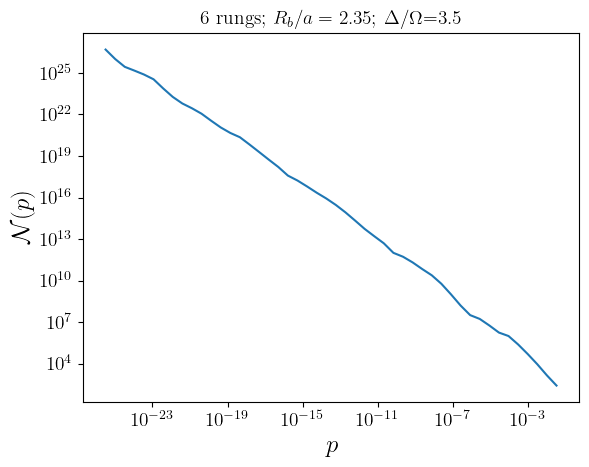}
\caption{\label{fig:density}Estimation of the density of probability $\npc$ for 6 rungs $R_b/a=2.35$ and $\Delta/\Omega$=3.5. We use 50 equally spaced intervals on a logarithmic scale between $p=10^{-26}$ and $10^{-1}$.}
\end{figure}

We can always use the integral representation
\beq
\label{eq:intrep}
\Sigma(\pla)=\int_0^{\pla}dp \mathcal{N}(p)p
\enq
by writing $\mathcal{N}(p)$ as a sum of delta functions weighted with possible degeneracies.
This notation becomes useful if for $p$ small enough and
in a large enough system, we can approximate $\npc$ by a continuous function and interpret this quantity as a density of states with a probability $p$. In other words,
$\npc \Delta p$ is approximately the number of states with a probability between $p-\dpc$ and $p+\dpc$.
If $\Sigma$ changes but not too rapidly over an interval $\Delta p$ we can try to use the continuous approximation
\beq
\Sigma(p+\dpc)-\Sigma(p-\dpc)\simeq \Delta p\npc p
\enq
Since there are only 4096 states for 6 rungs, $\Delta p $ cannot be too small
and the fluctuations observed for different choices of $\Delta p $ are not negligible.
A numerical example where we used
50 equally spaced intervals on a logarithmic scale between $p=10^{-26}$ and $10^{-1}$ is shown in Fig. \ref{fig:density}.

Fig. \ref{fig:density} suggests
to use a power behavior to characterize  the region of low probability
\beq
\npc \simeq Cp^{-1-\zeta}.
\enq
However, some curvature and local fluctuations are present and this is not a high quality fit. For completeness these fits provide small positive values for $\zeta$. Numerical values obtained are around 0.2 and depend on the choice of the fitting interval.
It should also be noted that Eq. (\ref{eq:intrep}) converges at the lower end for $\zeta<1$. In addition, under this assumption, we have the approximate asymptotic behavior
\beq
\Sigma (p)\sim p^{1-\zeta}.
\enq

\FloatBarrier
\section{\texorpdfstring{Validation of the M\textsubscript{3} Protocol with DMRG}{Validation of the M3 Protocol with DMRG}}
\label{app:dmrg_validation}

To verify that our implementation of the M\textsubscript{3} readout‐error mitigation is correct and unbiased by hardware effects, we perform the cumulative distribution and mutual‐information analyses using the DMRG “ground‐truth” samples \cite{corona_thesis}.
\begin{enumerate}
  \item Starting from the full $10^9$‐shot DMRG bitstring sample for each ladder size $\nr=6,8,10$,
        we \emph{simulate} readout noise by flipping each bit with probabilities
        $p_{0\to1}=0.01$ and $p_{1\to0}=0.08$.
  \item We then apply the identical M\textsubscript{3} mitigation procedure used on the Aquila data.
  \item Finally, we construct the empirical cumulative distribution in Eq. ~\ref{cum_sum} and the truncated mutual information in Eq. ~\ref{eq:mi} with sigmoid fit inflection markers exactly as in the main text.
\end{enumerate}
Figures~\ref{fig:6rung-pair-DMRG}–\ref{fig:10rung-pair-DMRG} 
and the associated Tables~\ref{tab:inflection_6rungs}-\ref{tab:inflection_10rungs}
show, for each ladder size, the \emph{pre‐} and \emph{post‐mitigation} curves (no hardware data), demonstrating that M\textsubscript{3} accurately restores the ideal DMRG statistics.
\subsection{Six‐rung validation.}
\begin{figure}[htbp]
\centering
    \includegraphics[width=0.9\linewidth, trim= 0 0 0 14, clip]{dmrg_cumulative_probs_error_and_mitigated_6_rungs_sample_size_4405.png}
    \includegraphics[width=0.9\linewidth, trim= 0 0 0 14, clip]{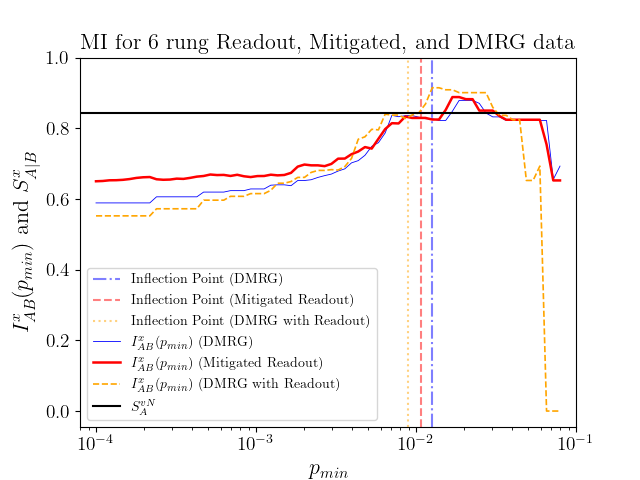}
  \caption{Comparison of cumulative distributions (top) and mutual information (bottom) for the 6-rung ladder. (No Hardware Data).}
  \label{fig:6rung-pair-DMRG}
\end{figure}

Figure~\ref{fig:6rung-pair-DMRG} shows that, for $\nr=6$, the mitigated cumulative distribution (green) lies almost exactly on top of the original DMRG (blue), demonstrating near‐perfect recovery of the true distribution after M\textsubscript{3}.  In the mutual information panel the mitigated and ideal curves are in good agreement, and even the purely “noisy” mutual information (yellow) tracks very closely for most truncation thresholds. This confirms both the correctness of our M\textsubscript{3} implementation and the surprising resilience of the mutual information estimator in small Hilbert spaces.

\begin{table}[htbp]
  \centering
  \begin{tabular}{|c|c|c|}
    \toprule
    Method & $p_{\min}^\star$ & $I_{AB}(p_{\min}^\star)$ \\
    \midrule
    Exact von Neumann entropy & — & 0.8441 \\
    DMRG inflection point       & $1.09(6)\times10^{-2}$ & 0.85(1) \\
    Mitigated inflection point   & $9.50\times10^{-3}$ & 0.84 \\
    Readout Simulated inflection   & $8.19\times10^{-3}$ & 0.85 \\
    \bottomrule
  \end{tabular}
  \caption{Inflection thresholds and corresponding mutual‐information values for the 6‐rung ladder.}
  \label{tab:inflection_6rungs}
\end{table}

\newpage

\FloatBarrier

\subsection{Eight‐rung validation.}
\begin{figure}[htbp]
\centering
    \includegraphics[width=0.9\linewidth]{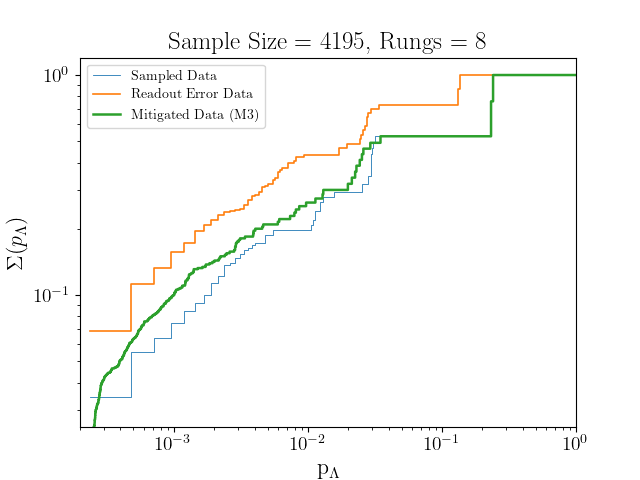}
    \caption{Comparison of cumulative distributions for the 8-rung ladder. (No Hardware Data)}
    \label{fig:8rung-cdf-DMRG}
\end{figure}

\begin{figure}
    \centering
    \includegraphics[width=0.9\linewidth, trim=0 0 0 14, clip]{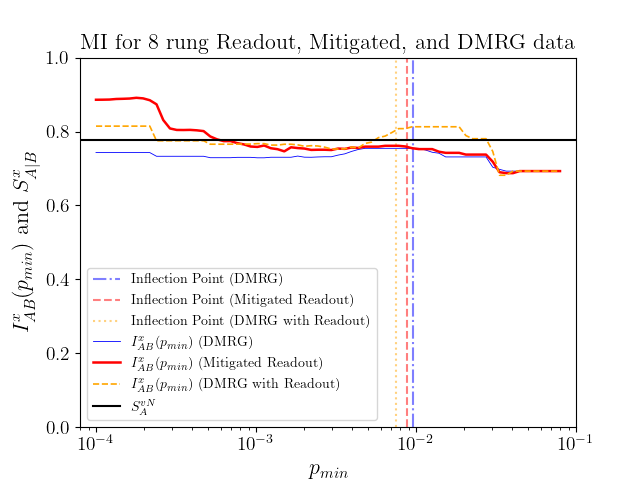}
      \caption{Comparison of mutual information for the 8-rung ladder. (No Hardware Data).}
      \label{fig:8rung-mi-DMRG}
\end{figure}

For $\nr=8$ (Figs~\ref{fig:8rung-cdf-DMRG} \& ~\ref{fig:8rung-mi-DMRG}), mitigation again restores the high‐$p$ region almost exactly, with the post‐M\textsubscript{3} cumulative distribution overlapping the ideal curve.  However, at the low‐probability end the mitigated curve rises slightly above DMRG, reflecting statistical limitations: our fixed‐size sample is now spread over a four‐times larger state space, so M\textsubscript{3} can only renormalize the \emph{observed} bitstrings.  In the mutual information panel, the mitigated and ideal curves coincide for large $p_{\min}$, and diverge only once rare events dominate; the simulated‐noise mutual information again remains a reasonable estimate.

\begin{table}[htbp]
  \centering
  \begin{tabular}{|c|c|c|}
    \toprule
    Method & $p_{\min}^\star$ & $I_{AB}(p_{\min}^\star)$ \\
    \midrule
    Exact von Neumann entropy & — & 0.7769 \\
    DMRG inflection point            & $1.01(8)\times10^{-2}$ & 0.749(4)   \\
    Mitigated inflection point   & $6.46\times10^{-3}$ & 0.75 \\
    Readout Simulated inflection   & $7.06\times10^{-3}$ & 0.81 \\
    \bottomrule
  \end{tabular}
  \caption{Inflection thresholds and corresponding mutual‐information values for the 8‐rung ladder.}
  \label{tab:inflection_8rungs}
\end{table}

\subsection{Ten‐rung validation.}
\begin{figure}[htbp]
\centering
    \includegraphics[width=0.9\linewidth, trim= 0 0 0 14, clip]{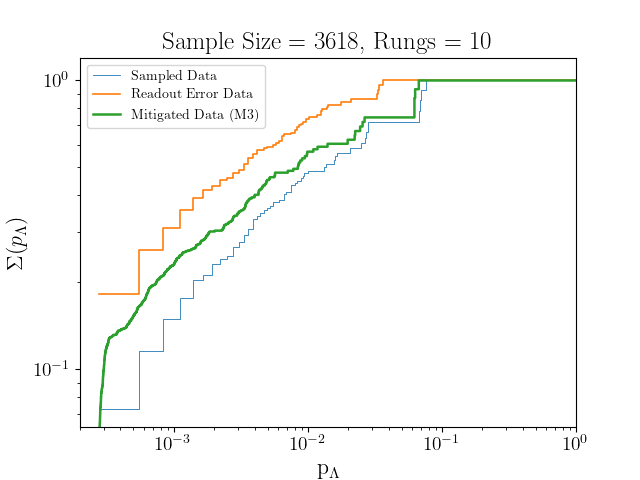}
    \includegraphics[width=0.9\linewidth, trim= 0 0 0 14, clip]{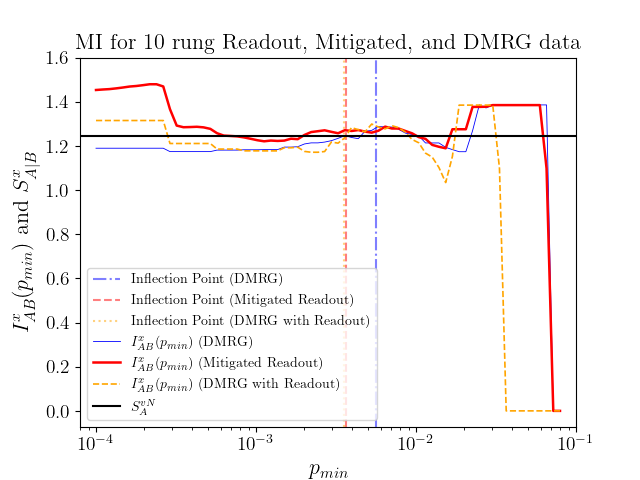}
  \caption{Comparison of cumulative distributions (top) and mutual information (bottom) for the 10-rung ladder. (No Hardware Data).}
  \label{fig:10rung-pair-DMRG}
\end{figure}

At $\nr=10$ (Fig.~\ref{fig:10rung-pair-DMRG}), the same trends intensify: the mitigated cumulative distribution recovers the top end almost perfectly but overshoots more strongly at small $p$, as the ratio of observed to total states shrinks further.  The mutual information panel shows excellent alignment between mitigated and ideal curves at high thresholds, with growing deviation only in the deep tails.  All three mutual information traces nonetheless bracket the exact entropy closely, underscoring that, even in large Hilbert spaces, the mutual information‐based truncation remains a robust approximation method.

\begin{table}[htbp]
  \centering
  \begin{tabular}{|c|c|c|}
    \toprule
    Method & $p_{\min}^\star$ & $I_{AB}(p_{\min}^\star)$ \\
    \midrule
    Exact von Neumann entropy       & —                  & 1.2455 \\
    DMRG inflection point       & $5.01(36)\times10^{-3}$ & 1.26(2) \\
    Mitigated inflection point     & $3.85\times10^{-3}$ & 1.23   \\
    Readout Simulated inflection    & $3.53\times10^{-3}$ & 1.20   \\
    \bottomrule
  \end{tabular}
  \caption{Inflection thresholds and corresponding mutual‐information values for the 10‐rung ladder. }
  \label{tab:inflection_10rungs}
\end{table}

\FloatBarrier
\section{Imperfect Adiabatic State Preparation}
\label{app:adiabatic}
To isolate errors arising from non-adiabatic evolution, we simulated the same time-dependent Hamiltonian Eq. (\ref{eq:genryd}), as used with Aquila, with the identical $4\,\mu$s ramp schedule (0→$\Omega_{\max}$→0, $-\Delta_{\max}$→$+\Delta_{\max}$) introduced in Sec.~\ref{mitigation_of_readout_error}.  We discretized the evolution into $20\,$ns steps and applied a first-order Trotter decomposition at each timestep.  Simultaneously, we computed the instantaneous ground state \(\ket{\psi_{\rm GS}(t)}\) by diagonalizing \(H(t)\).

\begin{figure}[htbp]
\centering
    \includegraphics[width=\linewidth]{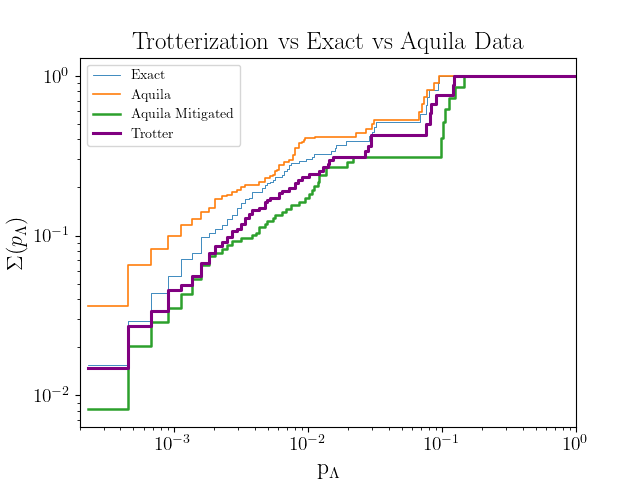}
    \caption{Final bitstring distributions for $\nr=6$: Exact Diagonalization ground state (blue), trotterized evolution (Purple), raw Aquila (orange), and M\textsubscript{3}-mitigated Aquila (green).  Even before readout, the Trotterized curve deviates significantly from the ideal exact diagonalization, indicating non-adiabatic transitions during the ramp.}
    \label{trotter_vs_exact_vs_aquila}
\end{figure}

Fig.~\ref{trotter_vs_exact_vs_aquila} compares the end-of-ramp probability distributions: the Trotterized state (purple) already differs markedly from the exact ground-state cumulative distribution (blue), mirroring much of the hardware’s raw and mitigated curves.  This confirms that imperfect adiabatic preparation—not just readout errors—accounts for a large fraction of the observed discrepancies.

\FloatBarrier
\section{Robustness to Ramp‐Schedule Variations}
\label{app:ramp_variation}

In order to test whether our overall conclusions depend sensitively on the exact adiabatic schedule, we repeated the Aquila runs with a slightly modified time series.  All other parameters remained unchanged (ladder sizes $\nr=6,8,10$, $R_b/a=2.35$, $a=4.1\,\mu$m, $\rho=2$, $C_6$, etc.), but the detuning hold was halved and extended:
See Table \ref{original_time_series} for the original schedule and Table \ref{modified_time_series} for the modified schedule.

\begin{table}[htbp]
    \centering
    \begin{tabular}{|c|c|c|c|c|}
    \hline
        $t$ & $0$ & $1\,\mu\mathrm s$ & $3.95\,\mu\mathrm s$ & $4\,\mu\mathrm s$ \\ \hline
        $\Omega(t)$ & $0$ & $\Omega_{\max}$ & $\Omega_{\max}$ & $0$ \\ \hline
        $\Delta(t)$ & $-\tfrac12\,\Delta_{\max}$ & $-\tfrac12\,\Delta_{\max}$ & $+\Delta_{\max}$ & $+\Delta_{\max}$ \\ \hline
    \end{tabular}
    \caption{$4 \mu s$ modified ramp schedule. }
    \label{modified_time_series}
\end{table}

\begin{figure}[htbp]
\centering
  \includegraphics[width=0.8\linewidth, trim= 0 0 0 14, clip]{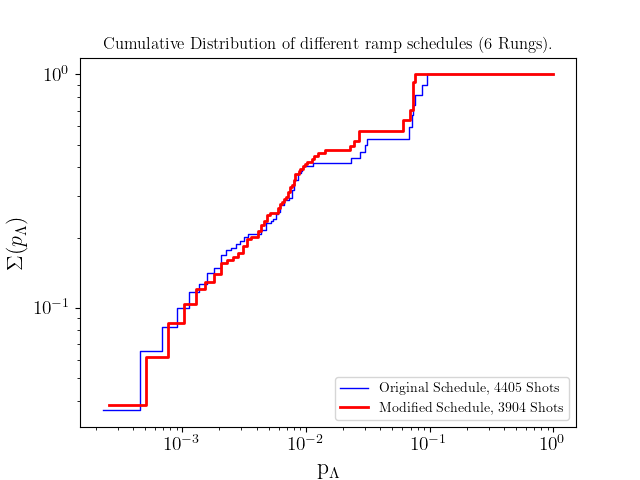}
  \caption{Cumulative Distributions for the 6-rung ladder under the two schedules above.}
  \label{fig:ramp-comparison6}
\end{figure}

Figures ~\ref{fig:ramp-comparison6} -~\ref{fig:ramp-comparison10} compare the final cumulative distributions of the two Aquila datasets.

\begin{figure}[htbp]
\centering
  \includegraphics[width=0.75\linewidth, trim= 0 0 0 14, clip]{6_rungs_aquila_avi_vs_asad.png}
  \caption{Cumulative Distributions for the 8-rung ladder under the two schedules above.}
  \label{fig:ramp-comparison8}
\end{figure}

\begin{figure}[htbp]
\centering
  \includegraphics[width=0.75\linewidth]{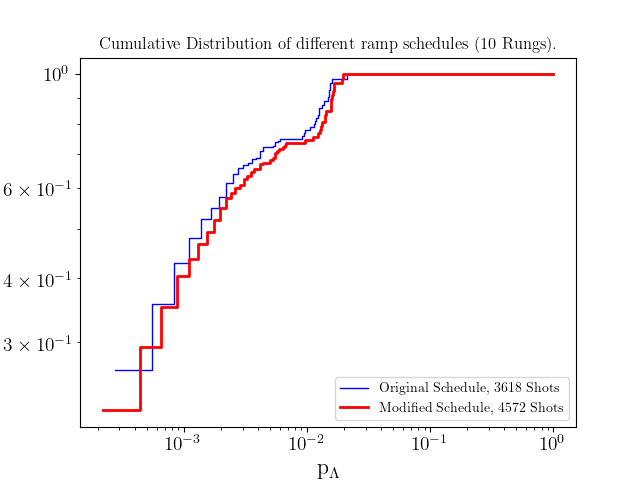}
  \caption{Empirical Cumulative Distributions for the 10‐rung ladder under the two schedules above.}
  \label{fig:ramp-comparison10}
\end{figure}
\FloatBarrier

The close overlap of the two cumulative distributions indicates that, despite the different ramp profiles, both $4 \mu s$ schedules adiabatically prepare essentially the same final state.

\FloatBarrier

\section{\texorpdfstring{12 $\mu$s Ramp Schedule}{12 us Ramp Schedule}}
\label{app:longer_schedule}
\begin{figure}[htbp]
\centering
  \includegraphics[width=0.85\linewidth, trim= 0 0 0 14, clip]{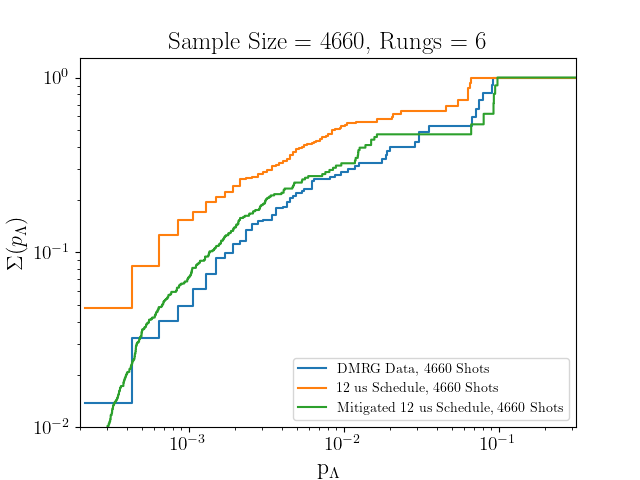}
  \caption{Empirical Cumulative Distributions for the 6‐rung ladder under the 12 $\mu s$ schedule above.}
  \label{fig:aquila_12us_6rungsA}
\end{figure}

\begin{figure}[htbp]
\centering
  \includegraphics[width=0.85\linewidth, trim= 0 0 0 14, clip]{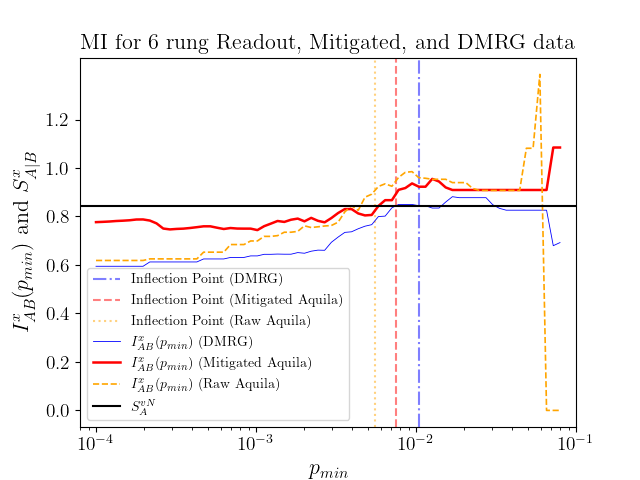}
  \caption{Empirical mutual information for the 6‐rung ladder under the 12 $\mu s$ schedule above.}
  \label{fig:aquila_12us_6rungsB}
\end{figure}

\begin{figure}[htbp]
\centering
  \includegraphics[width=0.81\linewidth, trim= 0 0 0 14, clip]{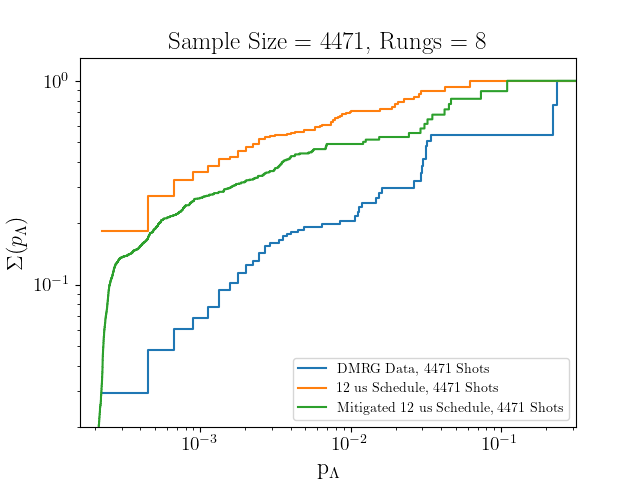}
  \includegraphics[width=0.81\linewidth, trim= 0 0 0 14, clip]{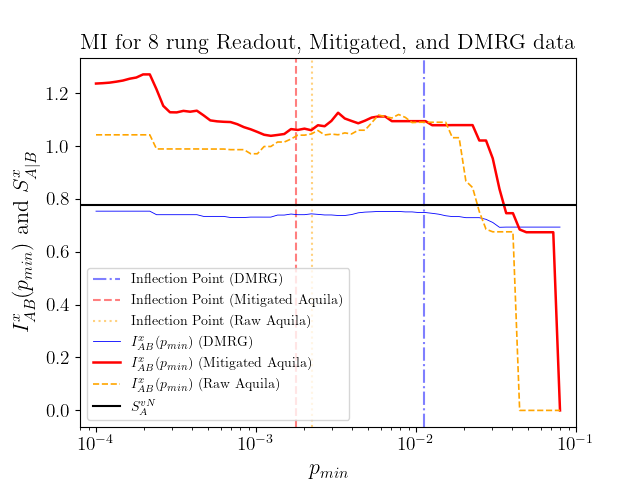}
  \caption{Empirical Cumulative Distributions and mutual information for the 8‐rung ladder under the 12 $\mu s$ schedule above.}
  \label{fig:aquila_12us_8rungs}
\end{figure}

\begin{figure}[htbp]
\centering
  \includegraphics[width=0.75\linewidth, trim= 0 0 0 14, clip]{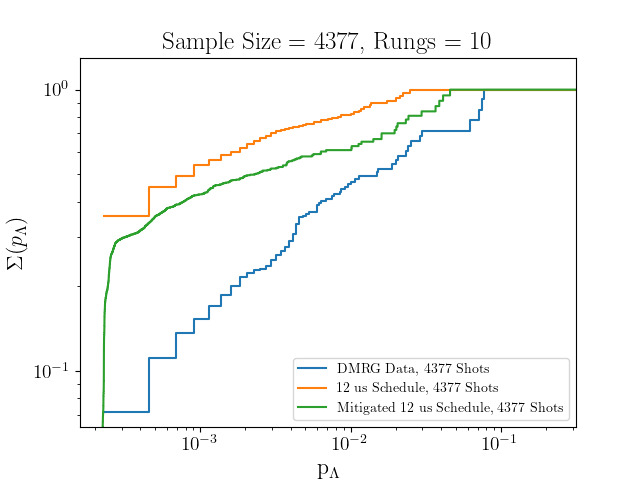}
  \caption{Empirical Cumulative Distributions for the 10‐rung ladder under the 12 $\mu s$ schedule above.}
  \label{fig:aquila_12us_10rungsA}
\end{figure}

\begin{figure}[htbp]
\centering
  \includegraphics[width=0.75\linewidth, trim= 0 0 0 14, clip]{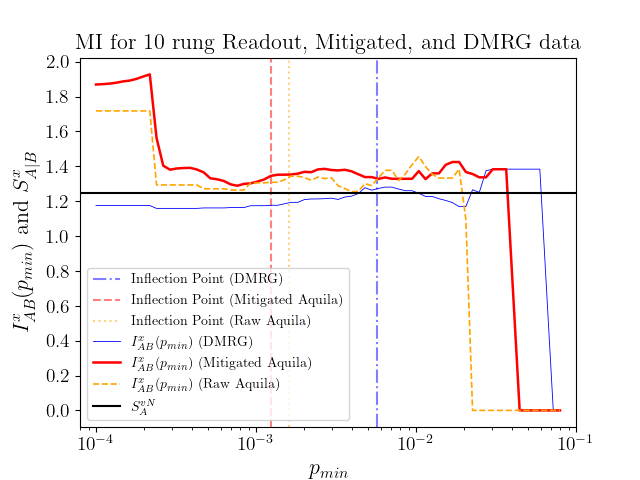}
  \caption{Empirical mutual information for the 10‐rung ladder under the 12 $\mu s$ schedule above.}
  \label{fig:aquila_12us_10rungsB}
\end{figure}

We now consider a longer schedule of  12 $\mu s$ with the time profiles given in Table \ref{tab:12usramp}.
\begin{table}[htbp]
\centering
    \begin{tabular}{|c|c|c|c|c|}
    \hline
    $t$ & $0$ & $0.5\,\mu\mathrm s$ & $11.95\,\mu\mathrm s$ & $12\,\mu\mathrm s$ \\ \hline
    $\Omega(t)$ & $0$ & $\Omega_{\max}$ & $\Omega_{\max}$ & $0$ \\
    $\Delta(t)$ & $-\Delta_{\max}$ & $-\Delta_{\max}$ & $+\Delta_{\max}$ & $+\Delta_{\max}$\\
    \hline
    \end{tabular}
  \caption{12 $\mu s$ ramp schedule.}
\label{tab:12usramp}
\end{table}
\FloatBarrier
Figures ~\ref{fig:aquila_12us_6rungsA} -~\ref{fig:aquila_12us_10rungsB} compare the final cumulative distributions of the two Aquila datasets. We observe that the improvement obtained numerically and shown in Figs. \ref{fig:4vstarget} \& \ref{fig:12vstarget} is not mirrored in the Aquila data for 12 $\mu s$ ramp-up schedules.

\subsection{6-rung ladder}

\newpage

\begin{table}[htbp]
  \centering
  \begin{tabular}{|c|c|c|}
    \toprule
    Method & $p_{\min}^\star$ & $I_{AB}(p_{\min}^\star)$ \\
    \midrule
    Exact von Neumann entropy & — & 0.8441 \\
    DMRG inflection point       & $1.09(6)\times10^{-2}$ & 0.85(1) \\
    Mitigated Aquila Inflection   & $7.48\times10^{-3}$ & 0.91 \\
    Raw Aquila inflection   & $5.58\times10^{-3}$ & 0.92 \\
    \bottomrule
  \end{tabular}
  \caption{Inflection thresholds and corresponding mutual‐information values for the 6‐rung ladder (12 us).}
  \label{tab:inflection_6rungs_12us}
\end{table}
\newpage

\subsection{8-rung ladder}
\FloatBarrier
\begin{table}[htbp]
  \centering
  \begin{tabular}{|c|c|c|}
    \toprule
    Method & $p_{\min}^\star$ & $I_{AB}(p_{\min}^\star)$ \\
    \midrule
    Exact von Neumann entropy & — & 0.7769 \\
    DMRG inflection point            & $1.01(8)\times10^{-2}$ & 0.749(4)   \\
    Mitigated Aquila Inflection   & $1.79\times10^{-3}$ & 1.06 \\
    Raw Aquila inflection   & $2.25\times10^{-3}$ & 1.04 \\
    \bottomrule
  \end{tabular}
  \caption{Inflection thresholds and corresponding mutual‐information values for the 8‐rung ladder (12 us).}
  \label{tab:inflection_8rungs_12us}
\end{table}

\newpage

\subsection{10-rung ladder}

\FloatBarrier

\begin{table}[htbp]
  \centering
  \begin{tabular}{|c|c|c|}
    \toprule
    Method & $p_{\min}^\star$ & $I_{AB}(p_{\min}^\star)$ \\
    \midrule
    Exact von Neumann entropy & — & 1.2455 \\
    DMRG inflection point       & $5.01(36)\times10^{-3}$ & 1.26(2) \\
    Mitigated Aquila Inflection   & $1.24\times10^{-3}$ & 1.34 \\
    Raw Aquila inflection   & $1.61\times10^{-3}$ & 1.34 \\
    \bottomrule
  \end{tabular}
  \caption{Inflection thresholds and corresponding mutual‐information values for the 10‐rung ladder (12 us).}
  \label{tab:inflection_10rungs_12us}
\end{table}

\FloatBarrier
\section{Multipartite Exploration}
\label{app:multipartite}
\FloatBarrier

\begin{figure}[htbp]
\centering
\includegraphics[width=0.98\linewidth]{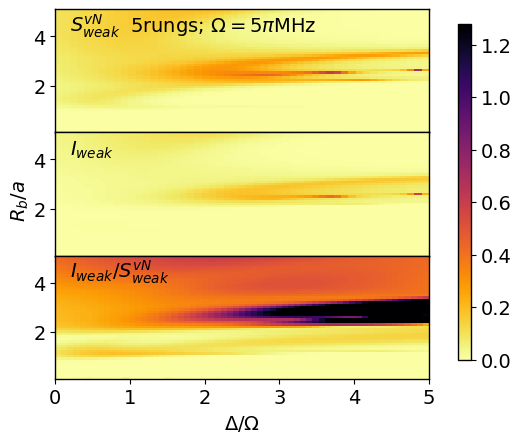}

    \caption{\label{fig:multipart_phase}5-rung Rydberg ladder, phase diagram in $\Delta/\Omega$ and $R_b/a$ showing the weak monotonicity. To the left is the quantity calculated for the von Neumann entanglement entropy, the mutual information, then their ratio.  
    Partitioning used:  DDAABBCCDD}
\end{figure}

In Sec.~\ref{sec:MI} in the main text we discuss a method for approximating the bipartite quantum entanglement with the classical mutual information. This method can be extended to look at a system partitioned into multiple parts.  We will test this by looking at a common inequality of information theory, weak monotonicity, and varying over a parameter.
We use the definition and the proposed approximation defined in equation \ref{eq:weak} for a system split into four parts $A$, $B$, $C$, and $D$.
\begin{equation}
    \begin{split}
    \label{eq:weak}
        S^{vN}_{weak} &= S^{vN}_{AB} + S^{vN}_{BC} - S^{vN}_A -S^{vN}_C \geq 0\\
        &\simeq I_{AB,CD}+I_{BC,AD}-I_{A,BCD}-I_{C,ABD}\\
        &=S^X_{AB}+S^X_{CD}+S^X_{BC}+S^X_{AD}\\
        &\text{ }-S^X_{A}-S^X_{BCD}-S^X_{C}-S^X_{ABD}, 
    \end{split}
\end{equation}
This quantity, together with the linear combinations of entropies that appear in strong subadditivity and related combinations, has been studied in the context of conformal field theory \cite{Hayden:2011ag,Lin:2023pvl}.   
This work has been taken forward to utilize such quantities to identify phase transitions \cite{ozz_multphase}.  Looking at Fig. \ref{fig:multipart_phase}, this is shown by the central lobes having higher entropy regions outlining them.  This peak is hypothesized to signify a phase transition.

Before looking at inherently multipartite entropy combinations like Eq. (\ref{eq:weak}) we must address how the calculation of the quantity looks with the mutual information lower bound. Because the mutual information always involves the Shannon entropy of a partition and its complement, there is some freedom in the notation used for the mutual information. Meaning, you could replace the mutual information quantities in $S^{vN}_{weak}$ with their complements. It would have been exactly the same to find the mutual information approximation by calculating $I^X_{CD,AB}+I^X_{AD,BC}-I^X_{BC,AD}-I^X_{AB,CD}$ in line two of Eq. (\ref{eq:weak}) instead.
\begin{figure}[htbp]
\centering
\includegraphics[width=0.90\linewidth]{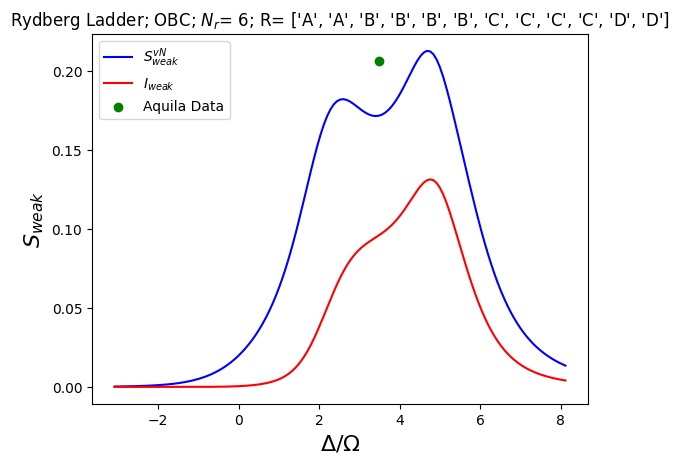}
\includegraphics[width=0.90\linewidth]{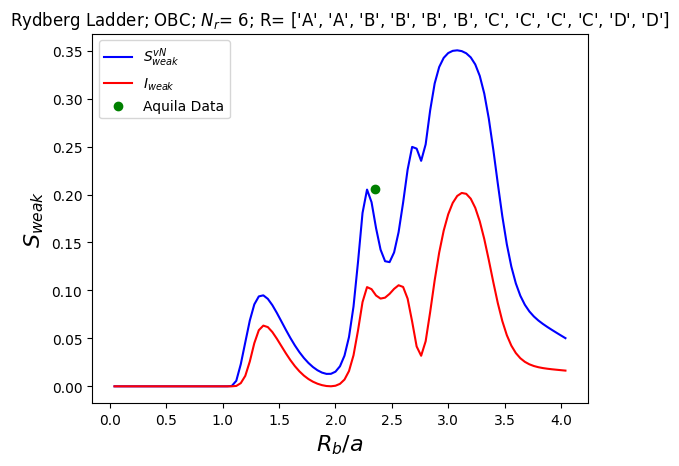}
\caption{\label{fig:multipart_ineqs}Six-rung Rydberg ladder, varying over $\Delta/\Omega$ with $R_b/a=2.35$ (top) and $R_b/a$ with $\Delta/\Omega=3.5$ (bottom) showing the weak monotonicity.}
\end{figure}

\FloatBarrier

Within Fig. \ref{fig:multipart_ineqs}, the mutual information is capturing the features of the von Neumann entanglement entropy. Additionally, approximately the same behavior is captured by each, specifically the peaking. We can observe that while the mutual information is a good approximation, it does not capture the exact features as well as the entanglement. We have also taken data from Aquila and found its $I^X_{weak}$ from mutual information of its probabilities on these graphs, and we find that the Aquila data is at least approximately close to the entanglement and mutual information for such parameters.
\begin{figure}[htbp]
\centering
    \includegraphics[width=\linewidth]{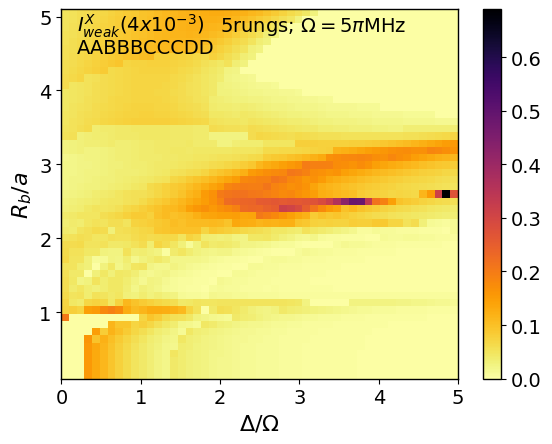}
    \caption{Mutual information for 5 rung Rydberg ladder of partitioning DDAABBCCDD filtered to 1/4000.}
    \label{fig:filtered_weakphase}
\end{figure}
\FloatBarrier

\begin{figure}[htbp]
\centering
    \includegraphics[width=0.9\linewidth]{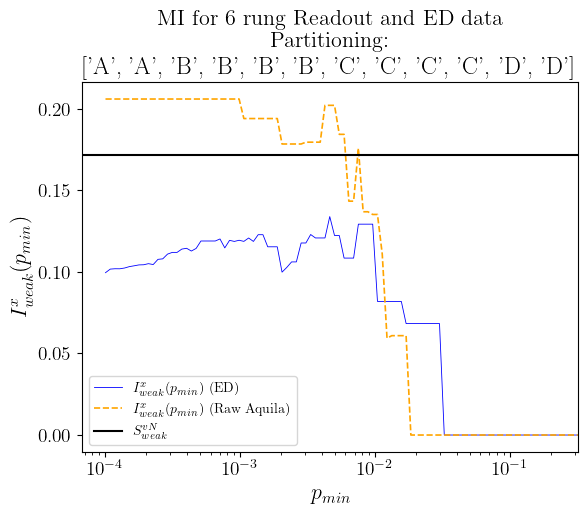}
    \caption{Applying a filtering scheme to the quantity in Eq. (\ref{eq:weak}) for computational and Aquila data with 6 rungs.}
    \label{fig:weak_filter6}
\end{figure}

\begin{figure}[htbp]
\centering
    \includegraphics[width=0.72\linewidth]{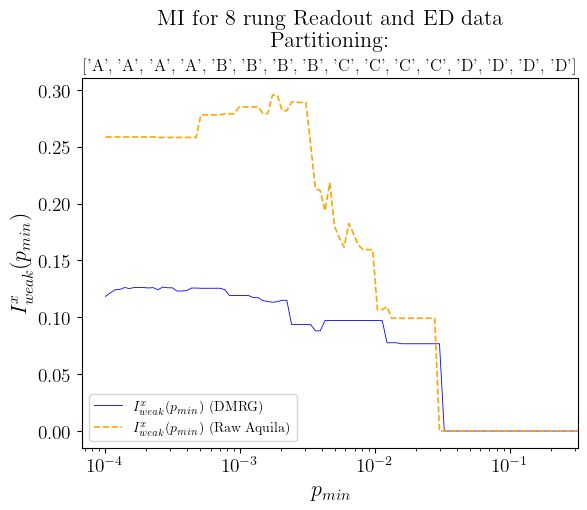}
    \caption{Applying a filtering scheme to the quantity in Eq. (\ref{eq:weak}) for computational and Aquila data with 8 rungs.}
    \label{fig:weak_filter8}
\end{figure}

Work is currently in progress \cite{ozz_multphase} to look further at using the inequality to identify critical points \cite{Lin:2023pvl}. We can further see this highlighted by the heat maps of phase structure in Figs. \ref{fig:multipart_phase}, \ref{fig:filtered_weakphase}, where the weak monotonicity's maximal locations well-define the phase regions. We take it a step further to look at the practical implications. By truncating our probabilities in the mutual information to a certain value, we can idealize the case of the mutual information given the returns from a quantum computer (Fig. \ref{fig:filtered_weakphase}). So our truncation to $1/4000$ is not dissimilar to a demonstration performed with $4000$ shots. The filtered plot is not perfect, but for the sections where the weak monotonicity peaking it is still clearly defined in comparison to the rest of the space even when filtered. Specifically, comparing to Fig. \ref{fig:multipart_phase}, we can see that much of the overall structure is still preserved after the filtering. This provides promising signs of an ability to utilize this method with the mutual information for identifying critical points with actual quantum computers.
This idea is further backed in Figs. \ref{fig:weak_filter6} \& \ref{fig:weak_filter8}, where we see the effects of different truncation values for the $S_{weak}$ quantity in ladders with different numbers of rungs. Moreover, this shows that given a sufficient shot total, one can attain an approximation to the quantity that will actually improve upon the mutual information bound, without exceeding the actual entanglement entropy.

\begin{figure}[htbp]
\centering
    \includegraphics[width=0.72\linewidth]{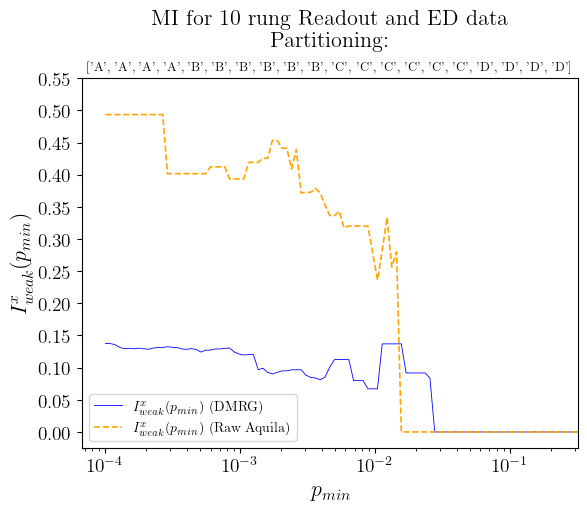}
    \caption{Applying the same filtering scheme for the 10-rung case.}
\end{figure}

Altogether, this provides promising signs that the mutual information approximation could be used as a scheme to identify phase transitions with bitstring results from a quantum computer, in place of the entanglement entropy \cite{ozz_multphase}.

\FloatBarrier

\end{document}